\documentclass[a4paper,11pt]{article} 
\usepackage[english]{babel}
\usepackage[stable]{footmisc}

\usepackage{amsfonts}
\usepackage{amsmath}
\numberwithin{equation}{section}

\usepackage{graphicx}
\usepackage[caption = true]{subfig}
\usepackage{caption}
\usepackage[colorlinks,bookmarks=true]{hyperref}
\usepackage{amsthm}
\usepackage{enumitem}
\usepackage{float}

\newtheorem{proposition}{Proposition}[section]

\newtheorem{corollary}{Corollary}[section]

\newcommand{\E}{\mathbb{E}}
\newcommand{\cF}{\mathcal{F}}

\newcommand{\p}{\partial}

\title{Volatility is rough}

\author{
Jim Gatheral\\Baruch College, City University of New York\\
jim.gatheral@baruch.cuny.edu\\$~~$\\
Thibault Jaisson\thanks{Thibault Jaisson gratefully acknowledges financial support from the chair ``Risques Financiers" of the Risk Foundation and the chair ``March\'es en Mutation" of the French Banking Federation.}\\ CMAP, \'Ecole Polytechnique Paris \\ thibault.jaisson@polytechnique.edu\\$~~$\\
Mathieu Rosenbaum\\ LPMA, Universit\'e Pierre et Marie Curie (Paris 6)\\
  mathieu.rosenbaum@upmc.fr}

\begin{document}

\maketitle

\vspace{-1mm}

\begin{abstract}

\noindent Estimating volatility from recent high frequency data, we revisit the question of the smoothness of the volatility process. Our main result is that log-volatility behaves essentially as a
fractional Brownian motion with Hurst exponent $H$ of order $0.1$, at any reasonable time scale.  This leads us to 
adopt the fractional stochastic volatility (FSV) model of Comte and Renault \cite{comte1998long}.  We call our model Rough FSV (RFSV) to underline that, in contrast to FSV, $H<1/2$.  We demonstrate that our RFSV model is remarkably consistent with financial time series data; one application is that it enables us to obtain improved forecasts of realized volatility.  Furthermore, we find that although volatility is not long memory in the RFSV model, classical statistical procedures aiming at detecting volatility persistence tend to conclude the presence of long memory in data generated from it. This sheds light on why long memory of volatility has been widely accepted as a stylized fact.  Finally, we provide a quantitative market microstructure-based foundation for our findings, relating the roughness of volatility to high frequency trading and order splitting.
%

%
\end{abstract}


\section{Introduction}





\subsection{Volatility modeling}

\noindent In the derivatives world, log-prices are often modeled as continuous semi-martingales. For a given asset with log-price $Y_t$, such a process takes the form
$$d Y_t=\mu_t dt+\sigma_t dW_t,$$
where $\mu_t$ is a drift term and $W_t$ is a one-dimensional Brownian motion. The term $\sigma_t$ denotes the volatility process and is the most important ingredient of the model. In the Black-Scholes framework, the volatility function is either constant or a deterministic function of time.  In Dupire's local volatility model, see \cite{dupire1994pricing}, the local volatility $\sigma(Y_t,t)$ is a deterministic function of the underlying price and time,  chosen to match observed European option prices exactly.  Such a model is by definition time-inhomogeneous;  its dynamics are highly unrealistic, typically generating future volatility surfaces (see Section \ref{sec:volSurfaceShape}  below) completely unlike those we observe.  A corollary of this is that prices of exotic options under local volatility can be substantially off-market.  On the other hand, in so-called stochastic volatility models, the volatility $\sigma_t$ is modeled as a continuous Brownian semi-martingale. Notable amongst such stochastic volatility models are the Hull and White model \cite{hull1993one}, the Heston model \cite{heston1993closed}, and the SABR model \cite{hagan2002managing}. Whilst stochastic volatility dynamics are more realistic than local volatility dynamics, generated option prices are not consistent with observed European option prices.  We refer to \cite{gatheral2006volatility} and \cite{musiela2006martingale} for more detailed reviews of the different approaches to volatility modeling. More recent market practice is to use local-stochastic-volatility (LSV) models which both fit the market exactly and generate reasonable dynamics.\\

\subsection{Fractional volatility}

\noindent In terms of the smoothness of the volatility process, the preceding models offer two possibilities: very regular sample paths in the case of Black-Scholes, and volatility trajectories with regularity close to that of Brownian motion for the local and stochastic volatility models. Starting from the stylized fact that volatility is a long memory process, various authors have proposed models that allow for a wider range of regularity for the volatility.  In a pioneering paper,  Comte and Renault \cite{comte1998long} proposed to model log-volatility using fractional Brownian motion (fBM for short), ensuring long memory by choosing the Hurst parameter $H>1/2$. A large literature has subsequently developed around such fractional volatility models, for example \cite{cheridito2003fractional,comte2012affine,rosenbaum2008estimation}. \\

\noindent The fBM $(W^H_t)_{t\in \mathbb{R}}$ with Hurst parameter $H\in (0,1)$, introduced in \cite{mandelbrot1968fractional}, is a centered self-similar Gaussian process with stationary increments satisfying for any $t\in \mathbb{R}$, $\Delta\geq 0$, $q>0$: 
\begin{equation}
\E[|W^H_{t+\Delta}-W^H_{t}|^q]=K_q\Delta^{qH},
\label{eq:fBM}
\end{equation}
with $K_q$ the moment of order $q$ of the absolute value of a standard Gaussian variable. For $H=1/2$, we retrieve the classical Brownian motion. The sample paths of $W^H$ are H\"older-continuous with exponent $r$, for any $r<H$\footnote{Actually $H$ corresponds to the regularity of the process in a more accurate way: in terms of Besov smoothness spaces, see  Section \ref{sec:howtomeas}.}. Finally, when $H>1/2$, the increments of the fBM are positively correlated and exhibit long memory in the sense that
$$\sum_{k=0}^{+\infty}\text{Cov}[W^H_1,W^H_k-W^H_{k-1}]=+\infty.$$ Indeed, $\text{Cov}[W^H_1,W^H_k-W^H_{k-1}]$ is of order $k^{2H-2}$ as $k \to \infty$. Note that in the case of the fBM, there is a one to one correspondence between regularity and long memory through the Hurst parameter $H$.\\

\noindent As mentioned earlier, the long memory property of the volatility process has been widely accepted as a stylized fact since the seminal analyses of Ding, Granger and Engle \cite{ding1993long}, Andersen and Bollerslev \cite{andersen1997intraday} and Andersen et al. \cite{andersen2001distribution}.  
Initially, it appears that the term {\it long memory} referred to the slow decay of the autocorrelation function (of absolute returns for example), anything slower than exponential. Over time however, it seems that this term has acquired the more precise meaning that the autocorrelation function is not integrable, see \cite{beran1994statistics}, and even more precisely that it decays as a power-law with exponent less than $1$. Much of the more recent literature, for example \cite{bentes2011stock,chen2006persistence,chronopoulou2011parameter}, assumes long memory in volatility in this more technical sense. Indeed, meaningful results can probably only be obtained under such a specification,  since it is not possible to estimate the asymptotic behavior of the covariance function without assuming a specific form. Nevertheless, analyses such as that of Andersen et al. \cite{andersen2001distribution} use data that predate the advent of high-frequency electronic trading, and the evidence for long memory has never been sufficient to satisfy remaining doubters such as Mikosch and St\u{a}ric\u{a} in \cite{mikosch2000really}.  To quote Rama Cont in \cite{Cont2007}:


\begin{quote}
... the econometric debate on the short range or long range nature of dependence in volatility still goes on (and may probably never be resolved)...
\end{quote}

\noindent One of our contributions in this paper is (we believe) to finally resolve this question, showing that the autocorrelation function of volatility does not behave as a power law, at least at usual time scales of observation.   This implies that when stated in term of the asymptotic behavior of the autocorrelation function, the long memory question can simply not be answered. Nevertheless, we are able to provide explicit expressions enabling us to analyze thoroughly the dependence structure of the volatility process. 



\subsection{The shape of the implied volatility surface}\label{sec:volSurfaceShape}

\begin{figure}[H]
\begin{center}
\includegraphics[width=1.1\linewidth]{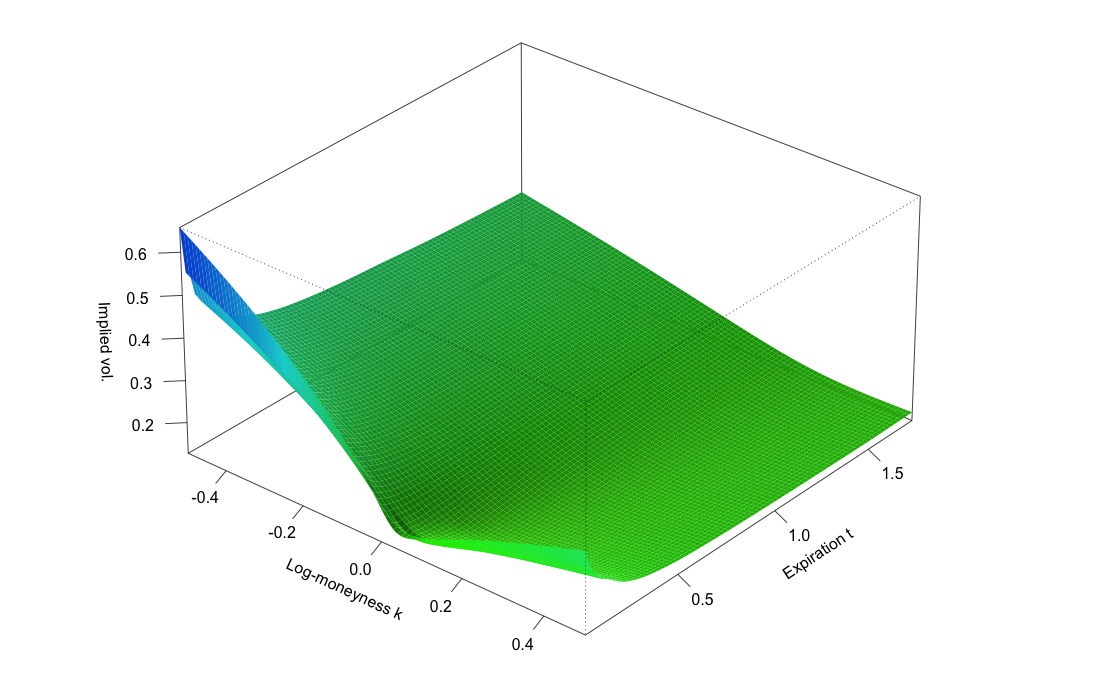}
\caption{The S\&P volatility surface as of June 20, 2013.}
\label{fig:volSurface}
\end{center}
\end{figure}

As is well-known, the implied volatility $\sigma_{\text{BS}}(k,\tau)$ of an option (with log-moneyness $k$ and time to expiration $\tau$) is the value of the volatility parameter in the Black-Scholes formula required to match the market price of that option.  Plotting implied volatility as a function of strike price and time to expiry generates the {\em volatility surface}, explored in detail in, for example, \cite{gatheral2006volatility}.  A typical such volatility surface generated from a ``stochastic volatility inspired'' (SVI) \cite{gatheral2014arbitrage} fit to closing SPX option prices as of June 20, 2013\footnote{Closing prices of SPX options for all available strikes and expirations as of June 20, 2013 were sourced from OptionMetrics (\url{www.optionmetrics.com}) via Wharton Research Data Services (WRDS).} is shown in Figure \ref{fig:volSurface}.   It is a stylized fact that, at least in equity markets, although the level and orientation of the volatility surface do change over time, the general overall shape of the volatility surface does not change, at least to a first approximation.  This suggests that it is desirable to model volatility as a time-homogenous process, {\em i.e.}  a process whose parameters are independent of price and time.  \\

\begin{figure}[H]
\begin{center}
\includegraphics[width=0.9\linewidth]{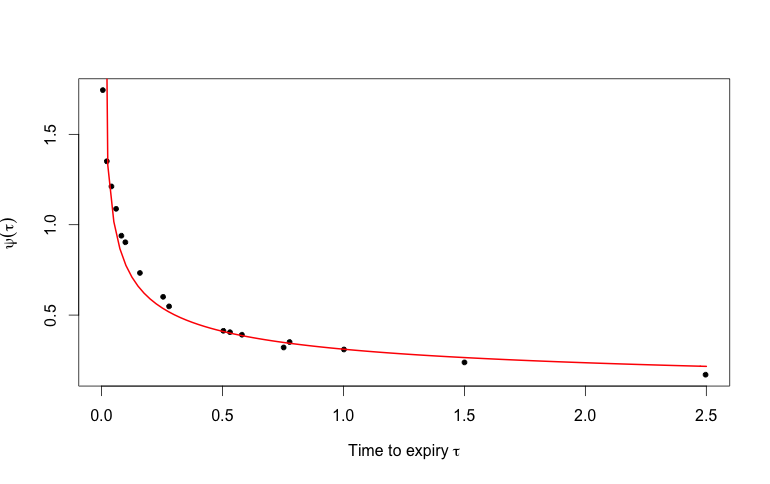}
\caption{The black dots are non-parametric estimates of the S\&P ATM volatility skews as of June 20, 2013; the red curve is the power-law fit $\psi(\tau) = A\,\tau^{-0.4}$.}
\label{fig:volSkew}
\end{center}
\end{figure}

\noindent However, conventional time-homogenous models of volatility such as the Hull and White, Heston, and SABR models do not fit the volatility surface.  In particular, as shown in Figure \ref{fig:volSkew}, the observed term structure of at-the-money ($k=0$) volatility skew
\[
\psi(\tau) :=\left|\frac{\p}{\p k}\sigma_{\text{BS}}(k,\tau)\right|_{k=0}
\]
is well-approximated by a power-law function of time to expiry $\tau$.  In contrast, conventional stochastic volatility models generate a term structure of at-the-money (ATM) skew that is {\em constant} for small $\tau$ and behaves as a sum of decaying exponentials for larger $\tau$.

\noindent In Section 3.3 of \cite{fukasawa2011asymptotic}, as an example of the application of his martingale expansion, Fukasawa shows that a stochastic volatility model where the volatility is driven by fractional Brownian motion with Hurst exponent $H$ generates an ATM volatility skew of the form $\psi(\tau) \sim \tau^{H-1/2}$,  at least for small $\tau$.  This is interesting in and of itself in that it provides a counterexample to the widespread belief that the explosion of the volatility smile as $\tau \to 0$ (as clearly seen in Figures \ref{fig:volSurface} and \ref{fig:volSkew}) implies the presence of jumps \cite{carr2003type}.   The main point here is that for a model of the sort analyzed by Fukasawa to generate a volatility surface with a reasonable shape, we would need to have a value of $H$ close to zero.  As we will see in Section \ref{sec:scaling}, our empirical estimates of $H$ from time series data are in fact very small.   \\

\noindent The volatility model that we will specify in Section \ref{secdefmodel}, driven by fBM with $H < 1/2$, therefore has the potential to be not only consistent with the empirically observed properties of the volatility time series but also consistent with the shape of the volatility surface.  In this paper, we focus on the modeling of the volatility time series.  A more detailed analysis of the consistency of our model with option prices is left for a future article.

\subsection{Main results and organization of the paper}
In Section \ref{sec:scaling}, we report our estimates of the smoothness of the log-volatility for selected assets. This smoothness parameter lies systematically between $0.08$ and $0.2$ (in the sense of H\"older regularity for example). Furthermore, we find that increments of the log-volatility are approximately normally distributed and that their moments enjoy a remarkable monofractal scaling property. This leads us to model the log of volatility using a fBM with Hurst parameter $H<1/2$ in Section \ref{sec:model}. Specifically we adopt the fractional stochastic volatility (FSV) model of Comte and Renault \cite{comte1998long}.  We call our model Rough FSV (RSFV) to underline that, in contrast to FSV, we take $H<1/2$.  We also show in the same section that the RFSV model is remarkably consistent with volatility time series data. The issue of volatility persistence is considered through the lens of the RFSV model in Section \ref{statistics}. Our main finding is that although the RFSV model does not have any long memory property, classical statistical procedures aiming at detecting volatility persistence tend to conclude the presence of long memory in data generated from it. This sheds new light on the supposed long memory in the volatility of financial data. In Section \ref{varianceswaps}, we apply our model to forecasting volatility. In particular, we show that RFSV volatility forecasts outperform conventional AR and HAR volatility forecasts. Finally, in Section \ref{hawkes}, we present a market microstructure explanation for the regularities we observe in the volatility process at the macroscopic scale.  We show that the empirical behavior of volatility may be explained in terms of order splitting and the high degree of endogeneity of the market ascribed to algorithmic trading. Some proofs are relegated to the appendix. 

\section{Smoothness of the volatility: empirical results}\label{sec:scaling}

In this section we report estimates of the smoothness of the volatility process for four assets: The DAX and Bund futures contracts, for which we estimate integrated variance directly from high frequency data using an estimator based on the model with uncertainty zones, \cite{robert2011new,robert2012volatility}, and the S\&P and NASDAQ indices, for which we use precomputed realized variance estimates from the Oxford-Man Institute of Quantitative Finance Realized Library\footnote{\url{http://realized.oxford-man.ox.ac.uk/data/download}. The Oxford-Man Institute's Realized Library contains a selection of daily non-parametric estimates of volatility of financial assets, including realized variance (rv) and realized kernel (rk) estimates.  A selection of such estimators is described and their performances compared in, for example, \cite{gatheral2010zero} .}.

\subsection{Estimating the smoothness of the volatility process}\label{sec:howtomeas}



\noindent  Let us first pretend that we have access to discrete observations of the volatility process, on a time grid with mesh $\Delta$ on $[0,T]$: $\sigma_0,~\sigma_{\Delta},\ldots,\sigma_{k\Delta}, \ldots,$ $k\in\{0,\lfloor T/\Delta\rfloor\}$. Set $N=\lfloor T/\Delta\rfloor$, then for $q\geq 0$, we define
$$m(q,\Delta)=\frac{1}{N}\sum_{k=1}^N|\log(\sigma_{k\Delta})-\log(\sigma_{(k-1)\Delta})|^q.$$
In the spirit of \cite{rosenbaum2011new}, our main assumption is that for some $s_q>0$ and $b_q>0$, as $\Delta$ tends to zero,
\begin{equation}\label{limiting}
N^{qs_q}m(q,\Delta)\rightarrow b_q.
\end{equation}
\noindent Under additional technical conditions, Equation \eqref{limiting} essentially says that the volatility process belongs to the Besov smoothness
space $\mathcal{B}^{s_q}_{q,\infty}$ and does not belong to $\mathcal{B}^{s_q'}_{q,\infty}$, for $s'_q>s_q$, see \cite{rosenbaum2009first}. Hence $s_q$ can really be viewed as the regularity of the volatility when measured in $l_q$ norm. In particular, functions in $\mathcal{B}^{s}_{q,\infty}$ for every $q>0$ enjoy the H\"older property with parameter $h$ for any $h<s$. For example, if $\log(\sigma_t)$ is a fBM with Hurst parameter $H$, then for any $q\geq 0$, Equation \eqref{limiting} holds in probability with $s_q=H$ and it can be shown that the sample paths of the process indeed belong to $\mathcal{B}^{H}_{q,\infty}$ almost surely. Assuming the increments of the log-volatility process are stationary and that a law of large number can be applied, $m(q,\Delta)$ can also be seen as the empirical counterpart of
$$\E[|\log(\sigma_{\Delta})-\log(\sigma_0)|^q].$$ 

\noindent Of course, the volatility process is not directly observable, and an exact computation of $m(q,\Delta)$ is not possible in practice. We must therefore proxy spot volatility values by appropriate estimated values. Since the minimal $\Delta$ will be equal to one day in the sequel, we proxy the (true) spot volatility daily at a fixed given time of the day (11 am for example).  Two daily spot volatility proxies will be considered:
\begin{itemize}

\item For our ultra high frequency intraday data (DAX future contracts and Bund future contracts\footnote{For every day, we only consider the future contract corresponding to the most liquid maturity.}, 1248 days from 13/05/2010 to 01/08/2014\footnote{Data kindly provided by QuantHouse EUROPE/ASIA, http://www.quanthouse.com.}), we use the estimator of the integrated variance from 10 am to 11 am London time obtained from the model with uncertainty zones, see \cite{robert2011new,robert2012volatility}. After renormalization, the resulting estimates of integrated variance over very short time intervals can be considered as good proxies for the unobservable spot variance. In particular, the one hour long window on which they are computed is small compared to the extra day time scales that will be of interest here. 

\item For the S\&P and NASDAQ indices\footnote{And also the CAC40, Nikkei and FTSE indices in some specific parts of the paper.}, we proxy daily spot variances by daily realized variance estimates from the Oxford-Man Institute of Quantitative Finance Realized Library (3,540 trading days from January 3, 2000 to March 31, 2014). Since these estimates of integrated variance are for the whole trading day, we expect estimates of the smoothness of the volatility process to be biased upwards, integration being a regularizing operation. We compute the extent of this bias by simulation in Section \ref{sec:simu}.

\end{itemize}

\noindent In the following, we retain the notation $m(q,\Delta)$ with the understanding that we are only proxying the (true) spot volatility as explained above. We now proceed to estimate the smoothness parameter $s_q$ for each $q$ by computing the $m(q,\Delta)$ for different values of $\Delta$ and regressing $\log m(q,\Delta)$ against $\log \Delta$. Note that for a given $\Delta$, several $m(q,\Delta)$ can be computed depending on the starting point. Our final measure of 
$m(q,\Delta)$ is the average of these values.

\subsection{DAX and Bund futures contracts}\label{sec:BundDax}


DAX and Bund futures are amongst the most liquid assets in the world and moreover, the model with uncertainty zones used to estimate volatility is known to apply well to them, see \cite{dayri2013large}. So we can be confident in the reliability of our volatility proxy. Nevertheless, as an extra check, we will confirm the quality of our volatility proxy by Monte Carlo simulation in Section \ref{sec:simu}.\\

\begin{figure}[H]
\begin{center}
\includegraphics[width=10cm,height=7cm]{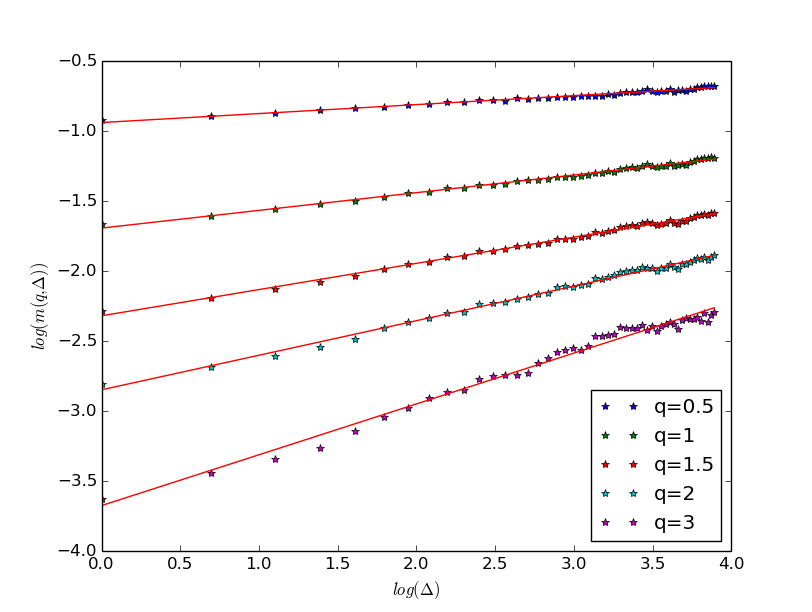}
\caption{$\log m(q,\Delta)$ as a function of $\log \Delta$, DAX.}
\label{scale_incr_log_voldax}
\end{center}
\end{figure}

\begin{figure}[H]
\begin{center}
\includegraphics[width=10cm,height=7cm]{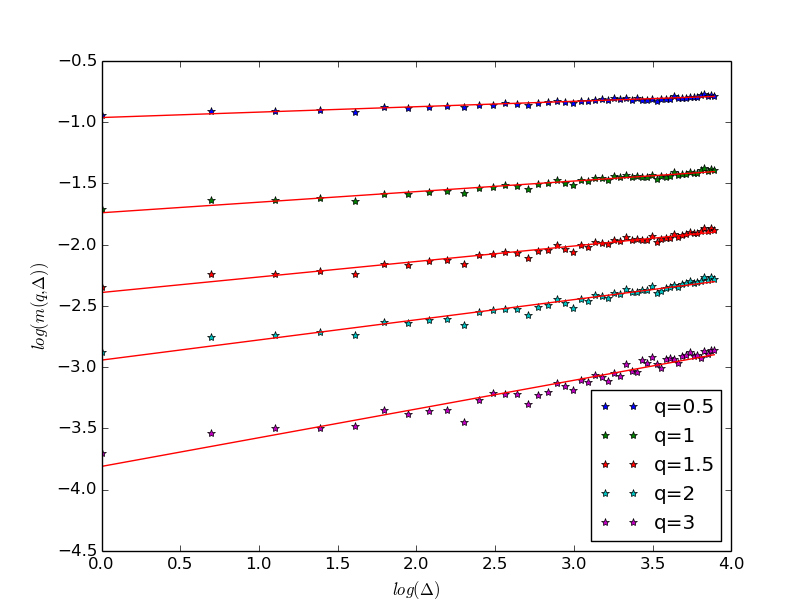}
\caption{$\log m(q,\Delta)$ as a function of $\log \Delta$, Bund.}
\label{scale_incr_log_volgbl}
\end{center}
\end{figure}

\noindent Plots of $\log m(q,\Delta)$ vs $\log \Delta $ for different values of $q$,  are displayed for the DAX in Figure \ref{scale_incr_log_voldax},  and for the Bund in Figure \ref{scale_incr_log_volgbl}.

\begin{figure}[H]
\begin{center}
\begin{tabular}{cc}
\includegraphics[width=7cm,height=7cm]{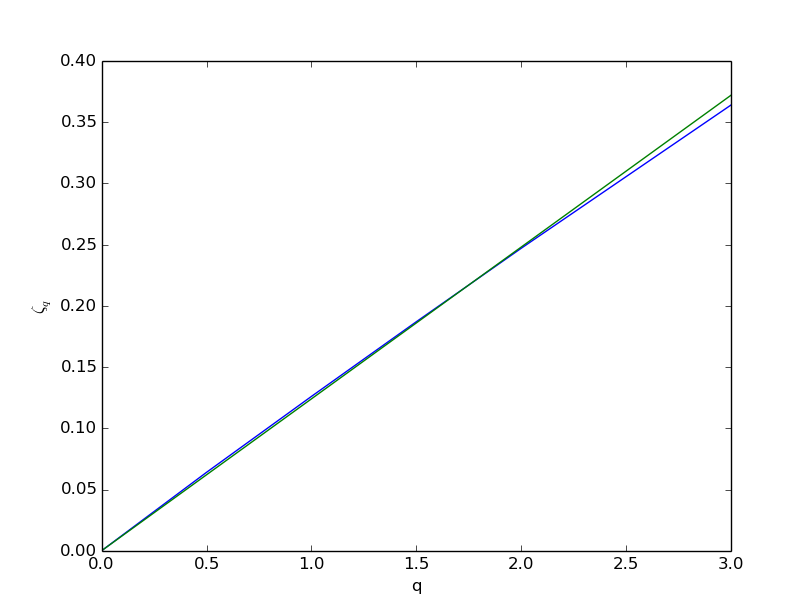}&\includegraphics[width=7cm,height=7cm]{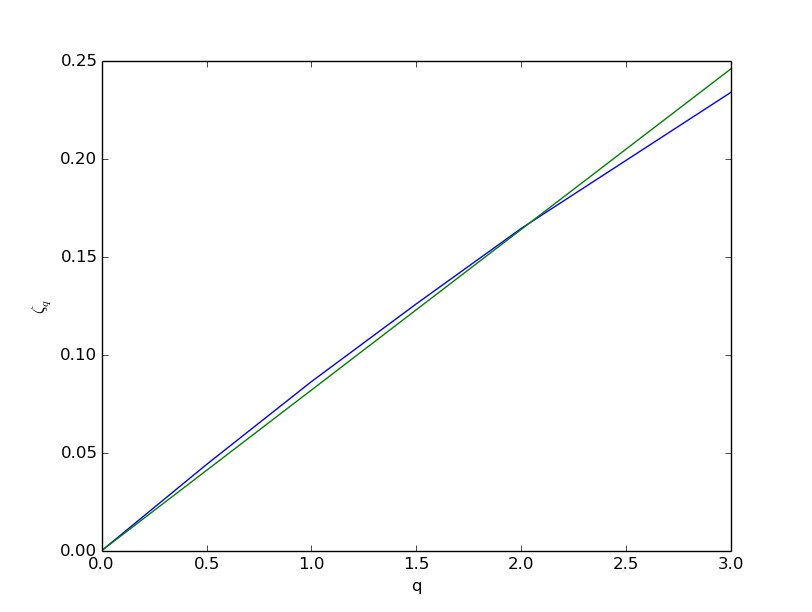}
\end{tabular}
\caption{$\zeta_q$ (blue) and $0.125\times q$ (green), DAX (left); $\zeta_q$ (blue) and $0.082\times q$ (green), Bund (right).}
\label{zeta_incr_log}
\end{center}
\end{figure}

\noindent For both DAX and Bund, for a given $q$, the points essentially lie on a straight line. Under stationarity assumptions, this implies that the log-volatility increments enjoy
the following scaling property in expectation:
$$\E[|\log(\sigma_{\Delta})-\log(\sigma_0)|^q]=K_q \Delta^{\zeta_q},$$ 
where $\zeta_q>0$ is the slope of the line associated to $q$. Moreover, the smoothness parameter $s_q$ does not seem to depend on $q$. Indeed, plotting $\zeta_q$ against $q$, we obtain that $\zeta_q\sim H\,q$ with $H$ equal to $0.125$ for the DAX and to $0.082$ for the Bund, see Figure \ref{zeta_incr_log}.

\noindent We remark that the graphs for $\zeta_q$ are actually very slightly concave.
However, we observe the same small concavity effect when we replace the log-volatility by simulations of a fBM with the same number of points.  We conclude that this effect relates to finite sample size and is thus not significant.

\subsection{S\&P and NASDAQ indices}\label{sec:Oxford}\label{sec:OxfordAnal}

\begin{figure}[H]
\begin{center}
\includegraphics[width=10cm,height=7cm]{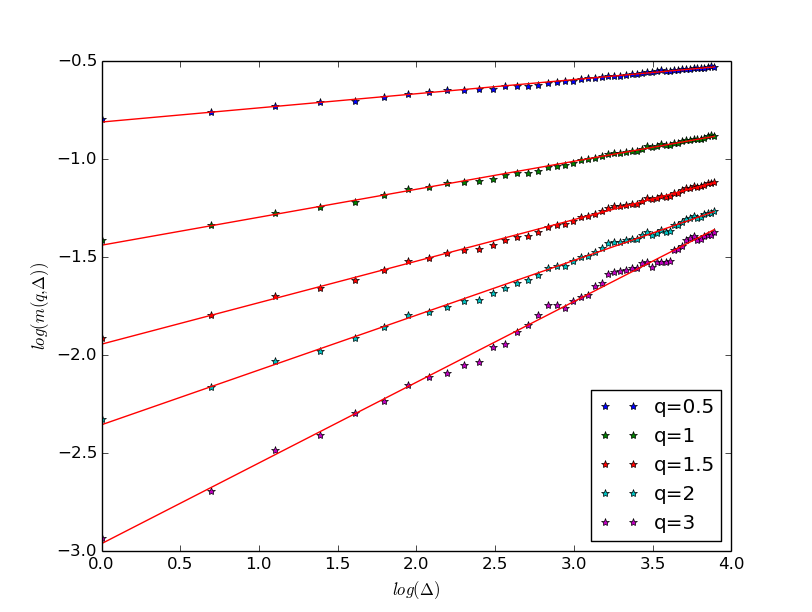}
\caption{$\log m(q,\Delta)$ as a function of $\log \Delta$, S\&P.}
\label{scale_incr_log_volsp}
\end{center}
\end{figure}

\begin{figure}[H]
\begin{center}
\includegraphics[width=10cm,height=7cm]{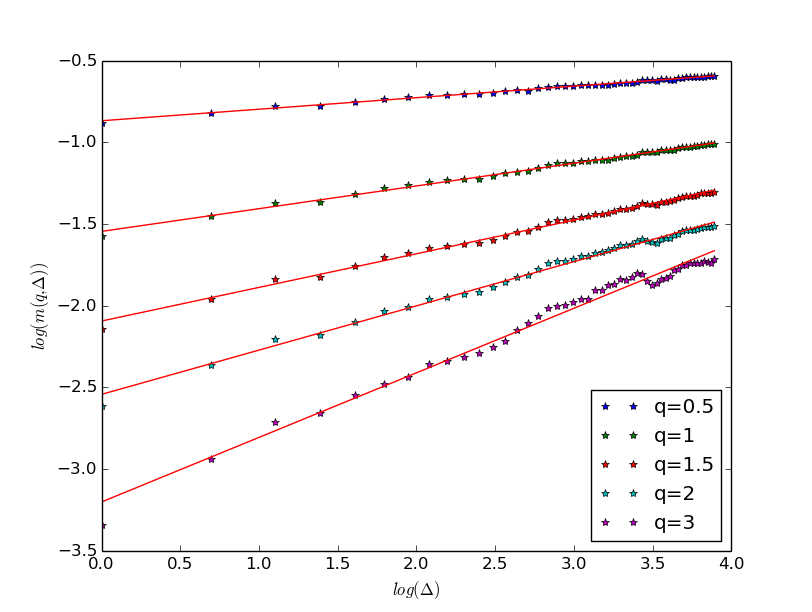}
\caption{$\log m(q,\Delta)$ as a function of $\log(\Delta)$, NASDAQ.}
\label{scale_incr_log_volNASDAQ}
\end{center}
\end{figure}

We report in Figure \ref{scale_incr_log_volsp} and Figure \ref{scale_incr_log_volNASDAQ} similar results for the S\&P and NASDAQ indices. The variance proxies used here are the precomputed 5-minute realized variance estimates for the whole trading day made publicly available by the Oxford-Man  Institute of Quantitative Finance.

\begin{figure}[H]
\begin{center}
\begin{tabular}{cc}
\includegraphics[width=7cm,height=7cm]{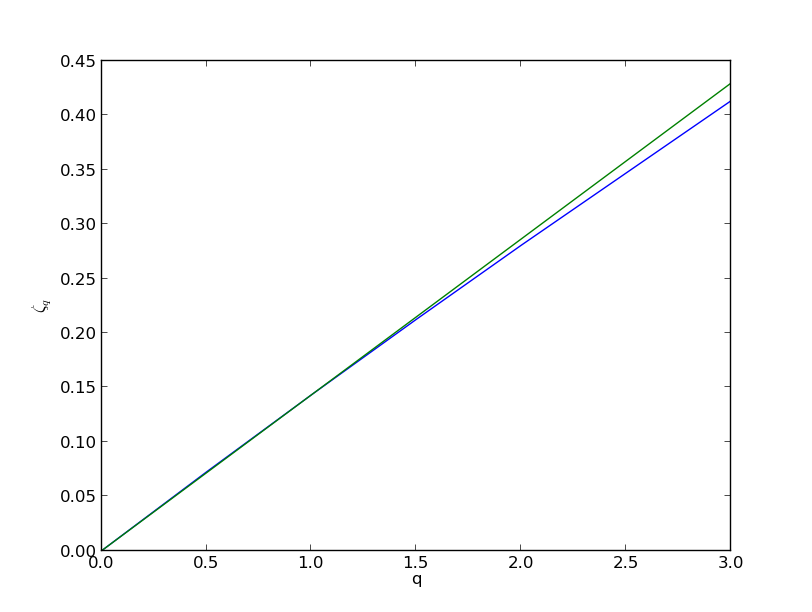}&\includegraphics[width=7cm,height=7cm]{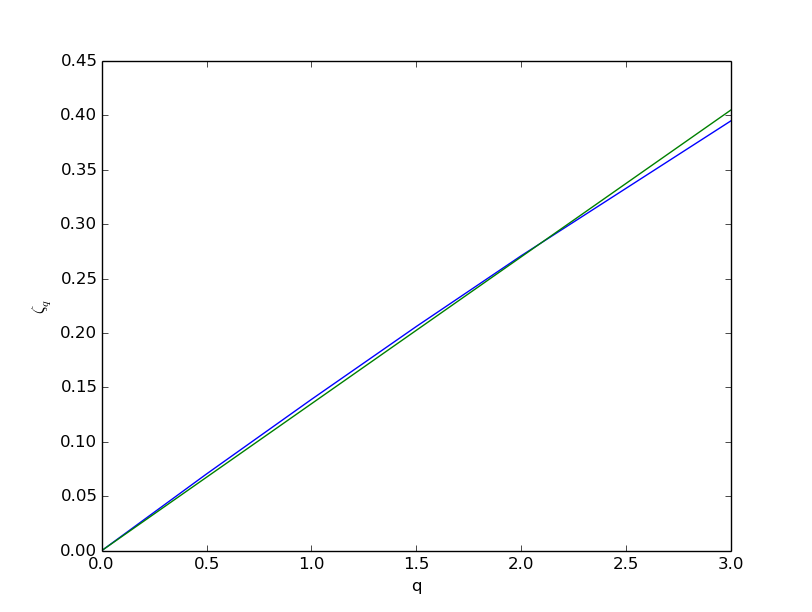}
\end{tabular}
\caption{$\zeta_q$ (blue) and $0.142\times q$ (green), S\&P (left); $\zeta_q$ (blue) and $0.139\times q$ (green), NASDAQ (right).}
\label{zeta_incr_log_sp_NASDAQ}
\end{center}
\end{figure}

\noindent We observe the same scaling property for the S\&P and NASDAQ indices as we observed for DAX and Bund futures and again, the $s_q$ do not depend on $q$. However, the estimated smoothnesses are slightly higher here: $H=0.142$ for the S\&P and $H=0.139$ for the NASDAQ, see Figure \ref{zeta_incr_log_sp_NASDAQ}. 

\noindent Once again, we do expect these smoothness estimates to be biased high because we are using whole-day realized variance estimates, as explained earlier in Section \ref{sec:scaling}.
 Finally, we remark that as for DAX and Bund futures, the graphs for $\zeta_q$ are slightly concave.

\subsection{Other indices}

Repeating the analysis of Section \ref{sec:Oxford} for each index in the Oxford-Man dataset, we find the $m(q,\Delta)$ present a universal scaling behavior. For each index and for $q=0.5,~1, ~1.5,~2,~3$, by doing a linear regression of $\log(m(q,\Delta))$ on $\log(\Delta)$ for $\Delta=1,...,30$, we obtain estimates of $\zeta_q$ that we summarize in Table \ref{table:OxfordSummary} in the appendix.








\subsection{Distribution of the increments of the log-volatility}

\begin{figure}[H]
\begin{center}
\subfloat[$\Delta=1$ day]{\includegraphics[width = 3in]{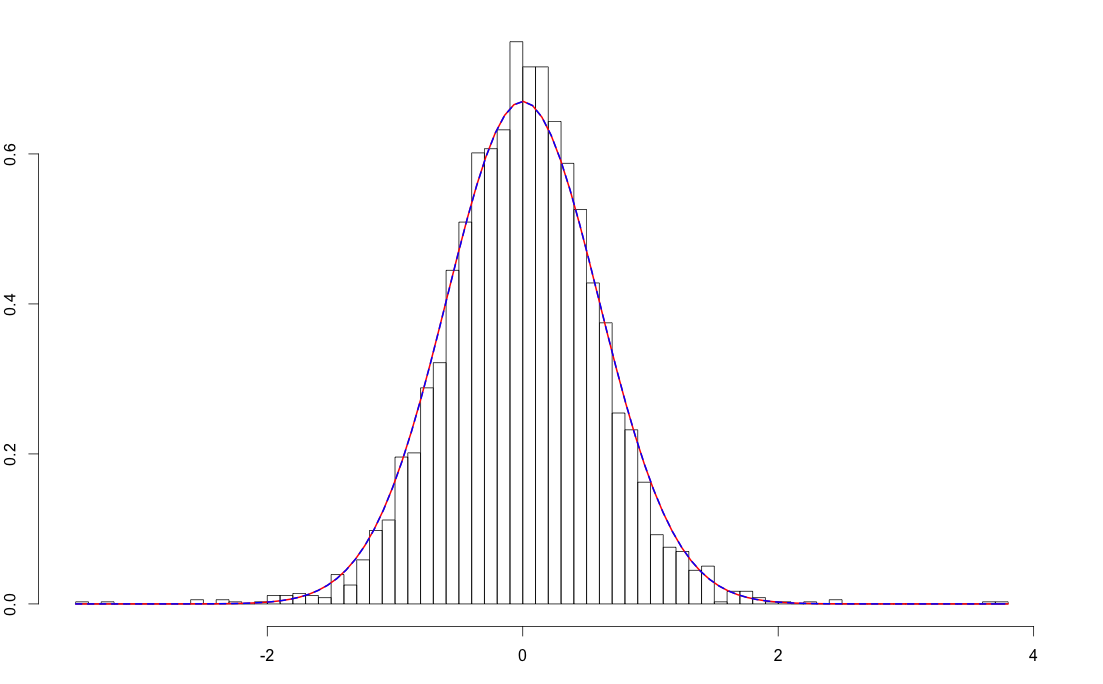}} 
\subfloat[$\Delta=5$ days]{\includegraphics[width = 3in]{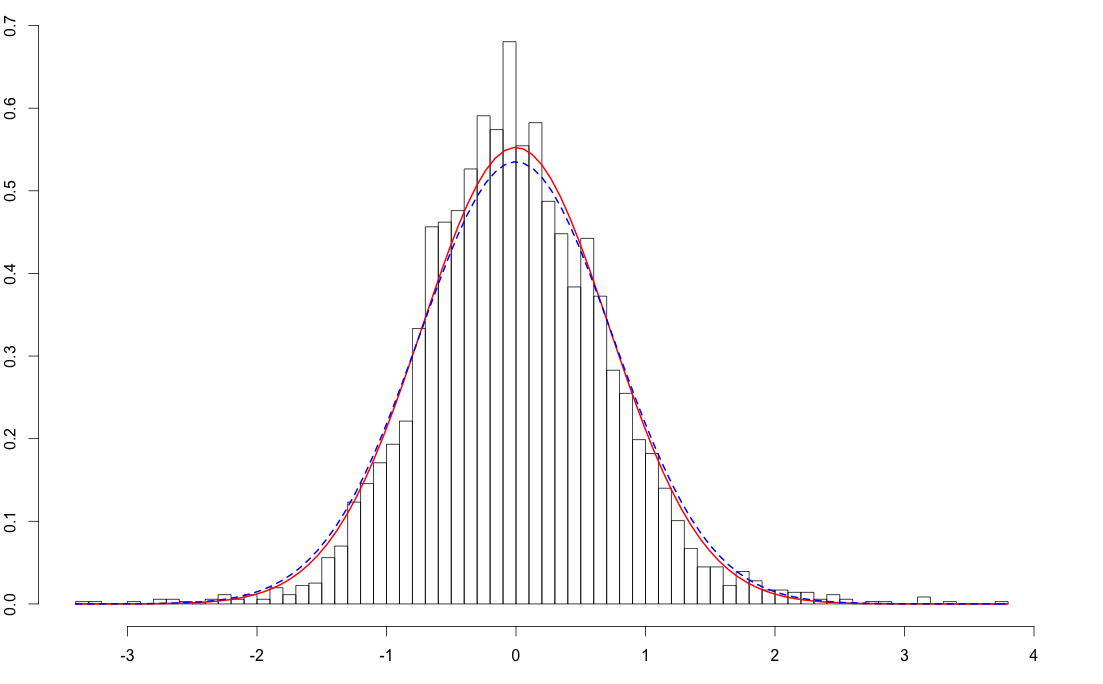}}\\
\subfloat[$\Delta=25$ days]{\includegraphics[width = 3in]{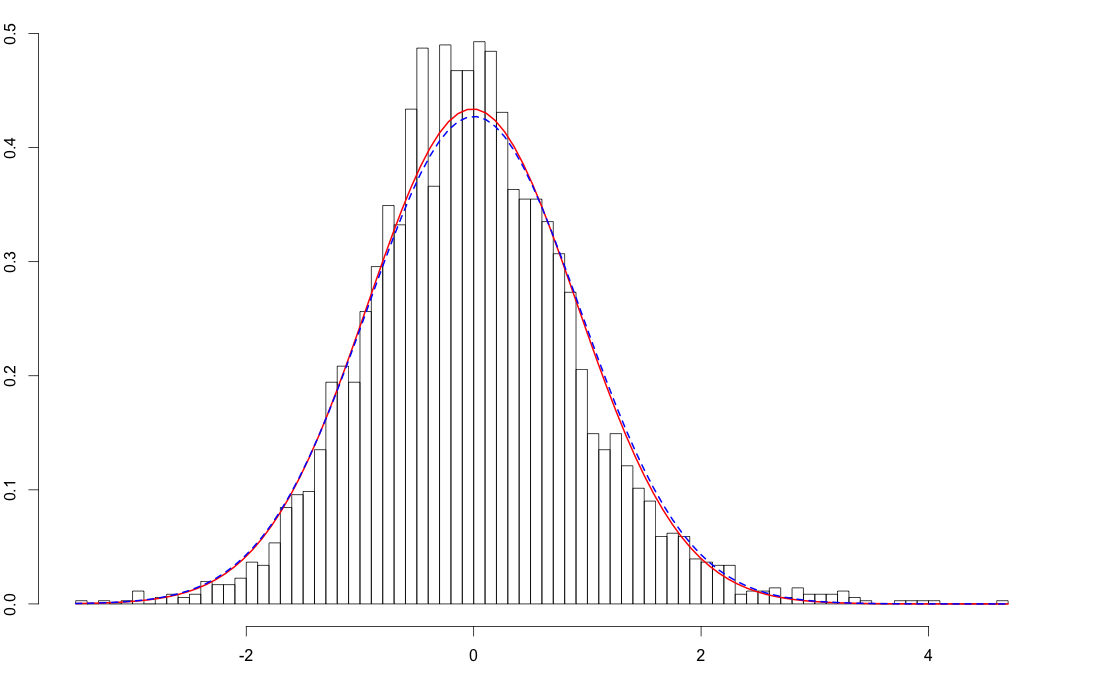}}
\subfloat[$\Delta=125$ days]{\includegraphics[width = 3in]{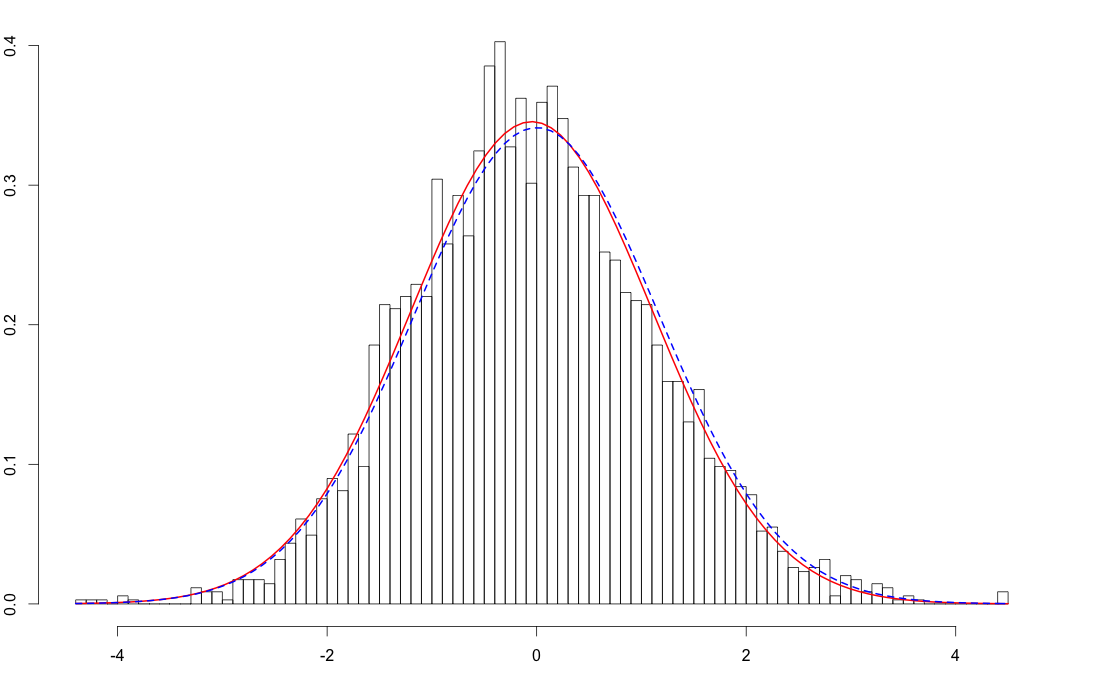}} 
\caption{Histograms for various lags $\Delta$ of the (overlapping) increments $\log \sigma_{t+\Delta}-\log \sigma_{t}$ of the S\&P log-volatility; normal fits in red; normal fit for $\Delta=1$ day rescaled by $\Delta^{H}$ in blue.}
\label{fig:SpxRvHist}
\end{center}
\end{figure}

Having established that all our underlying assets exhibit essentially the same scaling behavior\footnote{We have also verified that this scaling relationship holds for Crude Oil and Gold futures with similar smoothness estimates $\zeta_q$.}, we focus in the rest of the paper only on the S\&P index, unless specified otherwise.   That the distribution of increments of log-volatility is close to Gaussian is a well-established stylized fact reported for example in the papers \cite{andersen2001stock}  and \cite{andersen2001distribution} of Andersen et al.  Looking now at the histograms of the increments of the log-volatility in Figure \ref{fig:SpxRvHist} with the fitted normal density superimposed in red, we see that, for any $\Delta$, the empirical distributions of log-volatility increments are verified as being close to Gaussian.  More impressive still is that rescaling the $1$-day fit of the normal density by $\Delta^{H}$ generates (blue dashed) curves that are very close to the red fits of the normal density, consistent with the observed scaling.



%

\noindent The slight deviations from the Normal distribution observed in Figure \ref{fig:SpxRvHist} are again consistent with the computation of the empirical distribution of the increments of a fractional Brownian motion on a similar number of points.

\subsection{Does $H$ vary over time?}

In order to check whether our estimations of $H$ depends on the time interval, we split the Oxford-Man realized variance dataset into two halves and reestimate $H$ for each half separately. The results are presented in Table \ref{table:TimeVaryingH} in the appendix. We note that although the estimated $H$ all lie between $0.06$ and $0.20$, they seem to be higher in the second period which includes the financial crisis.

%
%

\section{A simple model compatible with the empirical smoothness of the volatility}

\label{sec:model}
In this section, we specify the Rough FSV model and demonstrate that it reproduces the empirical facts presented in Section \ref{sec:scaling}.

\subsection{Specification of the RFSV model}\label{secdefmodel}

In the previous section, we showed that, empirically, the increments of the log-volatility of various assets enjoy a scaling property with constant smoothness parameter and that their distribution is close to Gaussian. This naturally suggests the simple model:
\begin{equation}
\log \sigma_{t+\Delta} - \log \sigma_t =\nu\,\left( W^H_{t+\Delta}-W^H_t\right),
\label{eq:dataDriven}
\end{equation}
where $W^H$ is a fractional Brownian motion with Hurst parameter equal to the measured smoothness of the volatility and $\nu$ is a positive constant.  We may of course write \eqref{eq:dataDriven} under the form
\begin{equation}
\sigma_t=\sigma \exp\left\{{\nu\, W^H_t}\right\},
\label{eq:RFSV1}
\end{equation}
where $\sigma$ is another positive constant.
\\

%

\noindent 
However this model is not stationary, stationarity being desirable both for mathematical tractability and also to ensure reasonableness of the model at very large times. This leads us to impose stationarity by modeling the log-volatility as a fractional Ornstein-Uhlenbeck process (fOU process for short) with a very long reversion time scale.\\

\noindent  A stationary fOU process $(X_t)$ is defined as the stationary solution of the stochastic differential equation 
$$
dX_t=\nu\, dW^H_t -\alpha\,(X_t-m)dt,
$$
where $m \in \mathbb{R}$ and $\nu$ and $\alpha$ are positive parameters, see \cite{cheridito2003fractional}.
As for usual Ornstein-Uhlenbeck processes, there is an explicit form for the solution which is given by
\begin{equation}\label{eq:defOU}
X_t=\nu\int_{-\infty}^{t}e^{-\alpha(t-s)}dW^H_t +m.
\end{equation}
Here the stochastic integral with respect to fBM is simply a pathwise Riemann-Stieltjes integral, see again \cite{cheridito2003fractional}.\\

\noindent We thus arrive at the final specification of our Rough Fractional Stochastic Volatility (RFSV) model for the volatility on the time interval of interest $[0,T]$:
\begin{equation}\label{eq:RFSV}
\sigma_t=\exp\left\{X_t\right\},~t\in[0,T],
\end{equation}
where $(X_t)$ satisfies Equation \eqref{eq:defOU} for some $\nu>0,~\alpha>0$, $m\in \mathbb{R}$ and $H<1/2$ the measured smoothness of the volatility. Such a model is indeed stationary. However, if $\alpha \ll 1/T$, the log-volatility behaves locally (at time scales smaller than $T$) as a fBM. This observation is formalized in Proposition \ref{theoOUfBM} below.

\begin{proposition}\label{theoOUfBM}
Let $W^H$ be a fBM and $X^{\alpha}$ defined by \eqref{eq:defOU} for a given $\alpha>0$.  As $\alpha$ tends to zero, $$\E\Big[\underset{t\in[0,T]}{\emph{sup}}|X^\alpha_t-X^\alpha_0-\nu W^H_t|\Big]\rightarrow 0.$$ 
\end{proposition}

\noindent The proof is given in Appendix \ref{sec:proof3.1}.\\

\noindent Proposition \ref{theoOUfBM} implies that in the RFSV model, if $\alpha \ll 1/T$, and we confine ourselves to the interval $[0,T]$ of interest, we can proceed as if the the log-volatility process were a fBM. Indeed, simply setting $\alpha=0$ in \eqref{eq:defOU} gives (at least formally) $X_t-X_s = \nu(W^H_t-W^H_s)$ and we immediately recover our simple non-stationary fBM model \eqref{eq:dataDriven}.\\

\noindent The following corollary implies that the (exact) scaling property of the fBM is approximately reproduced by the fOU process when $\alpha$ is small. 

\begin{corollary}
\label{covfou}
Let $q>0,~t>0,~\Delta>0$. As $\alpha$ tends to zero, we have $$\E[|X^\alpha_{t+\Delta}-X^\alpha_t|^q]\rightarrow\nu^q \,K_q \,\Delta^{qH}.$$
\end{corollary}

\noindent The proof is given in Appendix \ref{sec:proofCor3.1}.

\subsubsection*{RFSV versus FSV}

\noindent We recognize our RFSV model \eqref{eq:RFSV} as a particular case of the classical FSV model of Comte and Renault \cite{comte1998long}. The key difference is that here we take $H<1/2$ and $\alpha \ll 1/T$, whereas to accommodate the assumption of long memory, Comte and Renault have to choose $H>1/2$.  The analysis of Fukasawa referred to earlier in Section \ref{sec:volSurfaceShape} implies in particular that if $H>1/2$, the volatility skew function $\psi(\tau)$ is {\em increasing} in time to expiration $\tau$ (at least for small $\tau$), which is obviously completely inconsistent with the approximately $1/\sqrt{\tau}$ skew term structure that is observed.  To generate a decreasing term structure of volatility skew for longer expirations, Comte and Renault are then forced to choose $\alpha \gg 1/T$.  Consequently, for very short expirations ($\tau \ll 1/\alpha$), models of the Comte and Renault type with $H>1/2$ still generate a term structure of volatility skew that is inconsistent with the observed one, as explained for example in Section 4 of \cite{comte2012affine}.\\

%
\noindent In contrast, the choice $H<1/2$ enables us to reproduce both the observed smoothness of the volatility process and generate a term structure of volatility skew in agreement with the observed one.  The choice $H<1/2$ is also consistent with what is improperly called mean reversion by practitioners, which is the fact that if volatility is unusually high, it tends to decline and if it is unusually low, it tends to increase. Finally, taking $\alpha$ very small implies that the dynamics of our process is close to that of a fBM, see Proposition \ref{theoOUfBM}. This last point is particularly important. Indeed, recall that at the time scales we are interested in, the important feature we have in mind is really this fBM like-behavior of the log-volatility. \\


\noindent We could no doubt have considered other stationary models satisfying Proposition \ref{theoOUfBM} and Corollary \ref{covfou}, where log-volatility behaves as a fBM at reasonable time scales; the choice of the fOU process is probably the simplest way to accommodate this local behavior together with the stationarity property.

\subsection{RFSV model autocovariance functions}


From Proposition \ref{theoOUfBM} and Corollary \ref{covfou}, we easily deduce the following corollary, where $o(1)$ tends to zero as $\alpha$ tends to zero.

\begin{corollary}\label{cor:RFSVautocovariance}
\label{covfunction} 
Let $q>0,~t>0,~\Delta>0$. As $\alpha$ tends to zero, $$\emph{Cov}[X^\alpha_t,X^\alpha_{t+\Delta}]= \emph{Var}[X^\alpha_t]-\frac12\,\nu^2\,\Delta^{2H}+o(1).$$
\end{corollary}

\begin{figure}[H]
\begin{center}
\includegraphics[width=10cm,height=7cm]{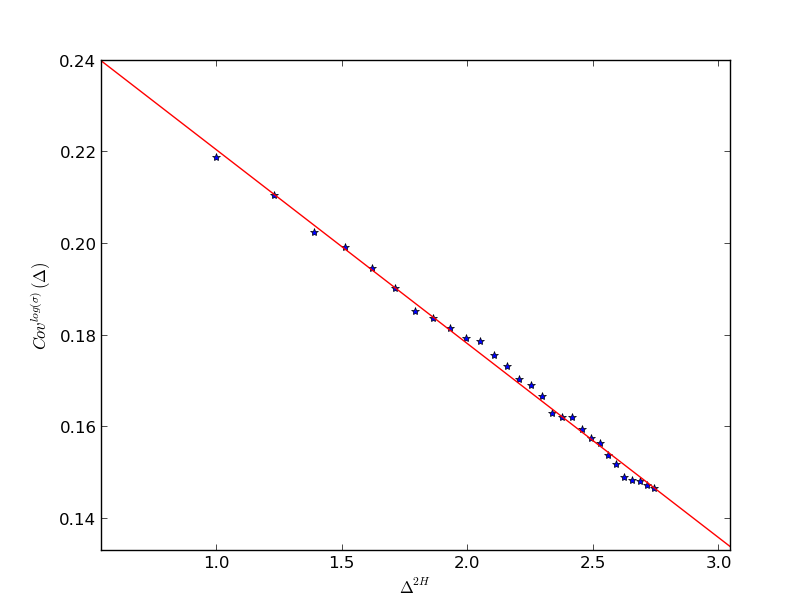}
\caption{Autocovariance of the log-volatility as a function of $\Delta^{2H}$ for $H=0.14$, S\&P.}
\label{cov_log_vol_pow}
\end{center}
\end{figure}

\noindent Consequently, in the RFSV model, for fixed $t$, the covariance between $X_t$ and $X_{t+\Delta}$ is linear with respect to $\Delta^{2H}$. This result is very well satisfied empirically. For example, in Figure \ref{cov_log_vol_pow}, we see that for the S\&P, the empirical autocovariance function of the log-volatility is indeed linear with respect to $\Delta^{2H}$. Note in passing that at the time scales we consider, the term $\text{Var}[X^\alpha_t]$ is higher than
$\frac{1}{2}\nu^2\,\Delta^{2H}$ in the expression for $\text{Cov}[X^\alpha_t,X^\alpha_{t+\Delta}]$.

\noindent Thanks to \cite{cheridito2003fractional}, we even have an exact formula for the autocovariance function of the log-volatility in the RFSV model:
\begin{eqnarray}
\label{eq:covfsvexact}
\text{Cov}[\log \sigma_t,\log \sigma_{t+\Delta}]&=& \frac{H\,(2\,H-1)\,\nu^2}{2\,\alpha^{2\,H}}\,\Bigg\{e^{-\alpha\,\Delta}\,\Gamma(2\,H-1)\nonumber \\
&+& e^{-\alpha\,\Delta}\,\int_0^{\alpha\,\Delta}\,\frac{e^u}{u^{2-2\,H}}\,du
+e^{\alpha\,\Delta}\,\int_{\alpha\,\Delta}^\infty\,\frac{e^{-u}}{u^{2-2\,H}}\,du\Bigg\},
\end{eqnarray}
and
\[
\text{Var}[\log \sigma_t] = \frac{H\,(2\,H-1)\,\nu^2}{\alpha^{2\,H}}\,\Gamma(2\,H-1),
\]
where $\Gamma$ denotes the Gamma function.\\

\begin{figure}[H]
\begin{center}
\includegraphics[width=10cm,height=7cm]{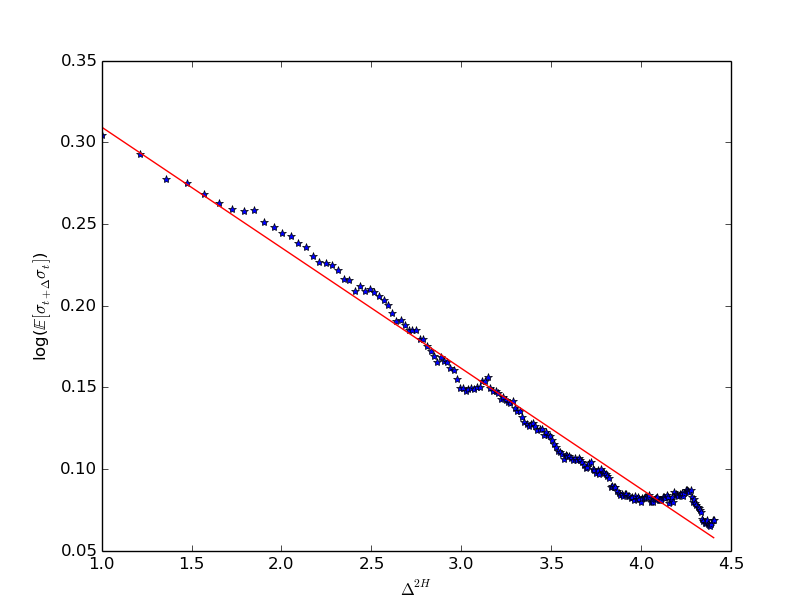}
\caption{Empirical counterpart of $\text{log}(\E[\sigma_{t+\Delta}\sigma_{t}])$ as a function of $\Delta^{2H}$, S\&P.}
\label{log_cov_vol_pow}
\end{center}
\end{figure}

\noindent Having computed the autocovariance function of the log-volatility, we now turn our attention to the volatility itself. We have
$$
\E[\sigma_{t+\Delta}\sigma_{t}]=\E[e^{X_t^{\alpha}+X^{\alpha}_{t+\Delta}}],
$$ 
with $X^{\alpha}$ defined by Equation \eqref{eq:defOU}.
Since $X^{\alpha}$ is a Gaussian process, we deduce that
$$
\E[\sigma_{t+\Delta}\sigma_{t}]= e^{\E[X^{\alpha}_t]+\E[X^{\alpha}_{t+\Delta}]+\text{Var}[X^{\alpha}_t]/2+
\text{Var}[X^{\alpha}_{t+\Delta}]/2+\text{Cov}[X^{\alpha}_t,X^{\alpha}_{t+\Delta}]}.
$$
Applying Corollary \ref{covfunction}, we obtain that when $\alpha$ is small,
$\E[\sigma_{t+\Delta}\sigma_{t}]$ is approximately equal to
\begin{equation}
e^{2\E[X^{\alpha}_t]+2\text{Var}[X^{\alpha}_t]}e^{-\nu^2\frac{\Delta^{2H}}{2}}.
\label{eq:volCovariance}
\end{equation}
It follows that in the RFSV model, $\log(\E[\sigma_{t+\Delta}\sigma_{t}])$ is also linear in $\Delta^{2H}$. This property is again very well satisfied on data, as shown by Figure \ref{log_cov_vol_pow}, where we plot the logarithm of the empirical counterpart of $\E[\sigma_{t+\Delta}\sigma_{t}]$ against $\Delta^{2H}$, for the S\&P with $H=0.14$.

\begin{figure}[H]
\begin{center}
\includegraphics[width=10cm,height=7cm]{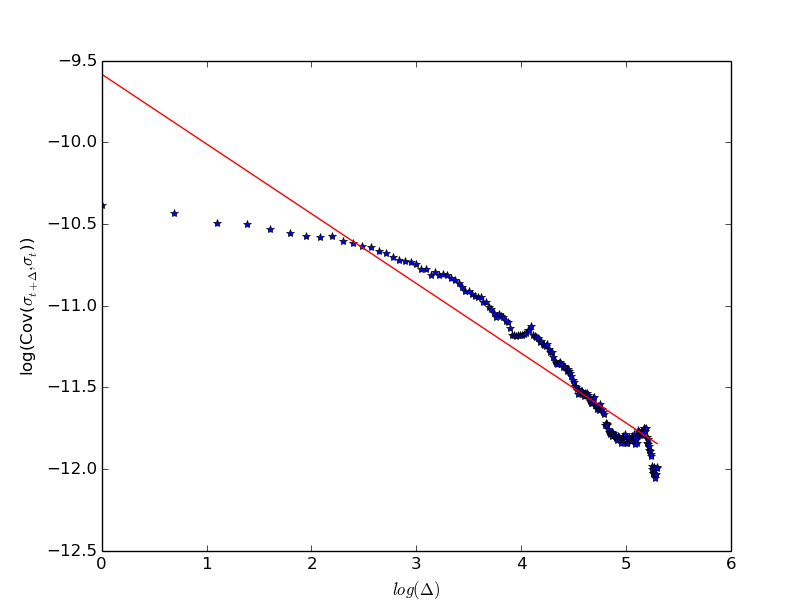}
\caption{Empirical counterpart of $\text{log}(\text{Cov}[\sigma_{t+\Delta},\sigma_{t}])$ as a function of $\text{log}(\Delta)$, S\&P.}
\label{log_cov_vol_log}
\end{center}
\end{figure}

\noindent We note that putting $\Delta^{2H}$ on the x-axis of Figure \ref{log_cov_vol_pow} is really crucial in order to retrieve linearity. In particular, a corollary of \eqref{eq:volCovariance} is that the autocovariance function of the volatility does not decay as a power law as widely believed; see Figure \ref{log_cov_vol_log} where we show that a log-log plot of the autocovariance function does not yield a straight line.

\subsection{RFSV versus FSV again}

To further demonstrate the incompatibility of the classical long memory FSV model with volatility data, consider the quantity $m(2,\Delta)$.  Recall that in the data (see Section \ref{sec:scaling}) we observe the linear relationship $\log m(2,\Delta) \approx \zeta_2\,\log \Delta + k$ for some constant $k$.
Also, in both FSV and RFSV, we can consider 
\begin{eqnarray*}
m(2,\Delta) &=& \E\left[(\log \sigma_{t+\Delta}-\log \sigma_t)^2\right]\\
&=& 2\,\left(\text{Var}[\log \sigma_t] - \text{Cov}[\log \sigma_t,\log \sigma_{t+\Delta}]\right).
\end{eqnarray*}
Therefore, using Equation \eqref{eq:covfsvexact}, we have a closed form formula for $m(2,\Delta)$.\\

\begin{figure}[H]
\begin{center}
\includegraphics[width = 0.9 \linewidth]{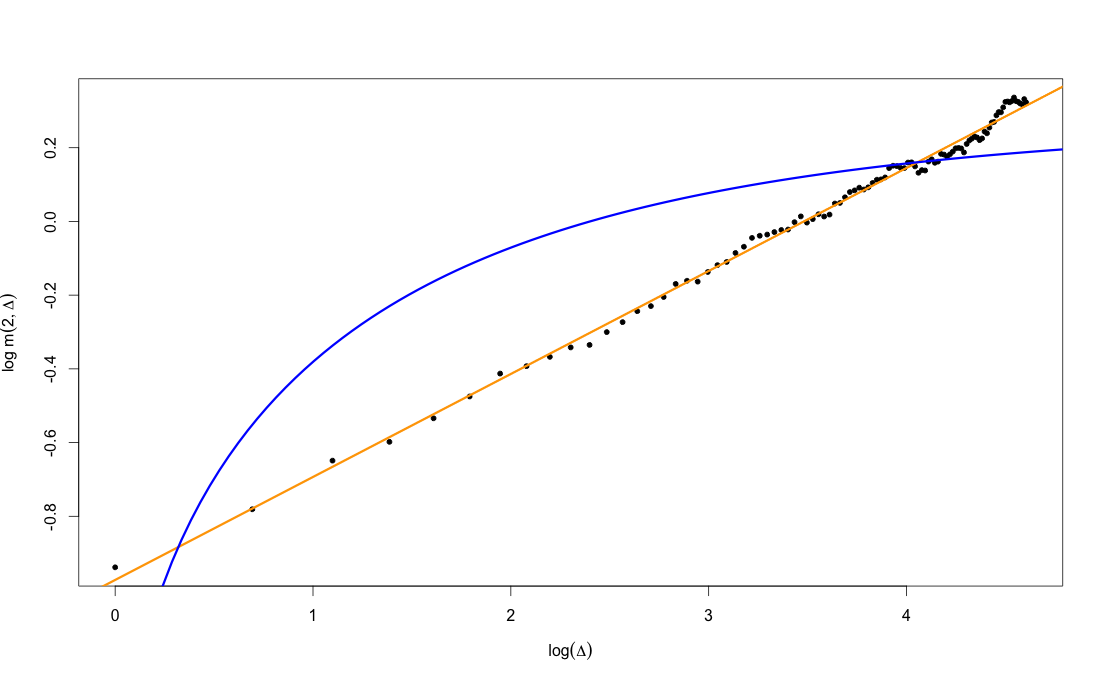}
\vspace{-10pt}
\caption{Long memory models such as the FSV model of Comte and Renault are not compatible with S\&P volatility data.  Black points are empirical estimates of $m(2,\Delta)$; the blue line is the FSV model with $\alpha=0.5$ and $H=0.53$; the orange line is the RFSV model with $\alpha=0$ and $H=0.14$.}
\label{fig:FSVvsFV}

\end{center}
\end{figure}

\noindent In Figure \ref{fig:FSVvsFV}, we plot $m(2,\Delta)$ with the parameters $H=0.53$, corresponding to the FSV model parameter estimate of Chronopoulou and Viens in \cite{chronopoulou2012estimation}, and $\alpha= 0.5$ to ensure some visible decay of the volatility skew.  The slope of $m(2,\Delta)$ in the FSV model for small lags is driven by the value of $H$; the lag at which $m(2,\Delta)$ begins to flatten and stationarity kicks in corresponds to a time scale of order $1/\alpha$.  It is clear from the picture that to fit the data, we must have $\alpha \ll 1/T$ and the value of $H$ must be set by the initial slope of the regression line, which as reported earlier in Section \ref{sec:scaling} is $\zeta_2=2\times 0.14$.

\subsection{Simulation-based analysis of the RFSV model} \label{sec:simu} 


\noindent Our goal in this section is to show that in terms of smoothness measures, one obtains on simulated data from the RFSV model the same behaviors as those observed on empirical data.  In particular, we would like to be able to quantify the positive bias associated with estimating $H$ from whole-day realized variance data as in Section \ref{sec:OxfordAnal} relative to using data from a one-hour window as in Section \ref{sec:BundDax}.\\

\noindent We simulate the RFSV model for $2,000$ days (chosen to be between the lengths of our two datasets). In order to account for the overnight effect, we simulate the volatility $\sigma_t$\footnote{To simulate the fBM, we use a spectral method with 40,000,000 points (20,000 points per day). We then simulate $X$ taking $X_{(n+1)\delta}-X_{n\delta}=\nu (W^H_{(n+1)\delta}-W^H_{n\delta})+\alpha \delta(m-X_{n\delta})$ (with $\delta=1/20 000$).} and efficient price $P_t$\footnote{$P_{(n+1)\delta}-P_{n\delta}=P_{n\delta}\sigma_{n\delta}\sqrt{\delta}\, U_n$ where the $U_n$ are iid standard Gaussian variables.}
over the whole day. The parameters: $H=0.14$, $\nu=0.3$, $m=X_0=-5$ and $\alpha=5 \times 10^{-4}$, are chosen to be consistent with our empirical estimates from Section \ref{sec:scaling}. 
To model microstructure effects such as the discreteness of the price grid, we consider that the observed price process is generated from $P_t$ using the uncertainty zones model of  \cite{robert2011new} with tick value $5\times 10^{-4}$ and parameter $\eta=0.25$.\\

\noindent Exactly as  in Section \ref{sec:scaling}, for each of the 2,000 days, we consider two volatility proxies obtained from the observed price and based on:
\begin{itemize}
\item The integrated variance estimator using the model with uncertainty zones over one hour windows, from 10 am to 11 am.
\item The 5 minutes realized variance estimator, over eight hours windows (the trading day). 
\end{itemize}

\begin{figure}[H]
\begin{center}
\includegraphics[width=0.75\linewidth]{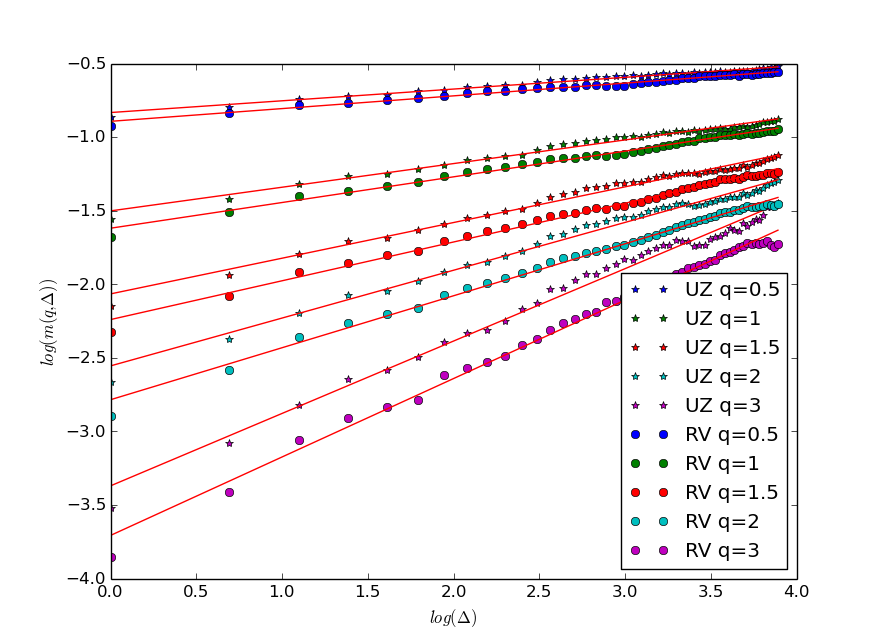}
\caption{$\log(m(q,\Delta))$ as a function of $\log(\Delta)$, simulated data, with realized variance and uncertainty zones estimators.}
\label{scale_incr_log_vol_simu_1h-8h}
\end{center}
\end{figure}

\noindent We now repeat our analysis of Section \ref{sec:scaling}, generating graphs analogous to Figures \ref{scale_incr_log_voldax}, \ref{scale_incr_log_volgbl}, \ref{scale_incr_log_volsp} and \ref{scale_incr_log_volNASDAQ} obtained on empirical data. Figure \ref{scale_incr_log_vol_simu_1h-8h} compares smoothness measures obtained using the uncertainty zones estimator on one-hour windows with those obtained using the realized variance estimator on 8-hour windows.

%
%


\noindent When the uncertainty zones estimator is applied on a one-hour window ($1/24$ of a simulated day)  as in Section \ref{sec:BundDax}, we estimate $H=0.16$, which is close to the true value $H=0.14$ used in the simulation.   
The results obtained with the realized variance estimator over daily eight-hour windows ($1/3$ of a simulated day) do exhibit the same scaling properties that we see in the empirical data with a smoothness parameter that does not depend on $q$. However, the estimated $H$ is biased slightly higher at around $0.18$. As discussed in Section \ref{sec:howtomeas}, this extra positive bias is no surprise and is due to the regularizing effect of the integral operator over the longer window.  We note also that the estimated values of $\nu$ (``volatility of volatility'' in some sense) obtained from the intercepts of the regressions, are lower with the longer time windows, again as expected.  A detailed computation of the bias in the estimated  $H$ associated with the choice of window length in an analogous but more tractable model is presented in Appendix \ref{sec:fSS}.\\

\begin{figure}[H]
\begin{center}
\includegraphics[width=15cm,height=11cm]{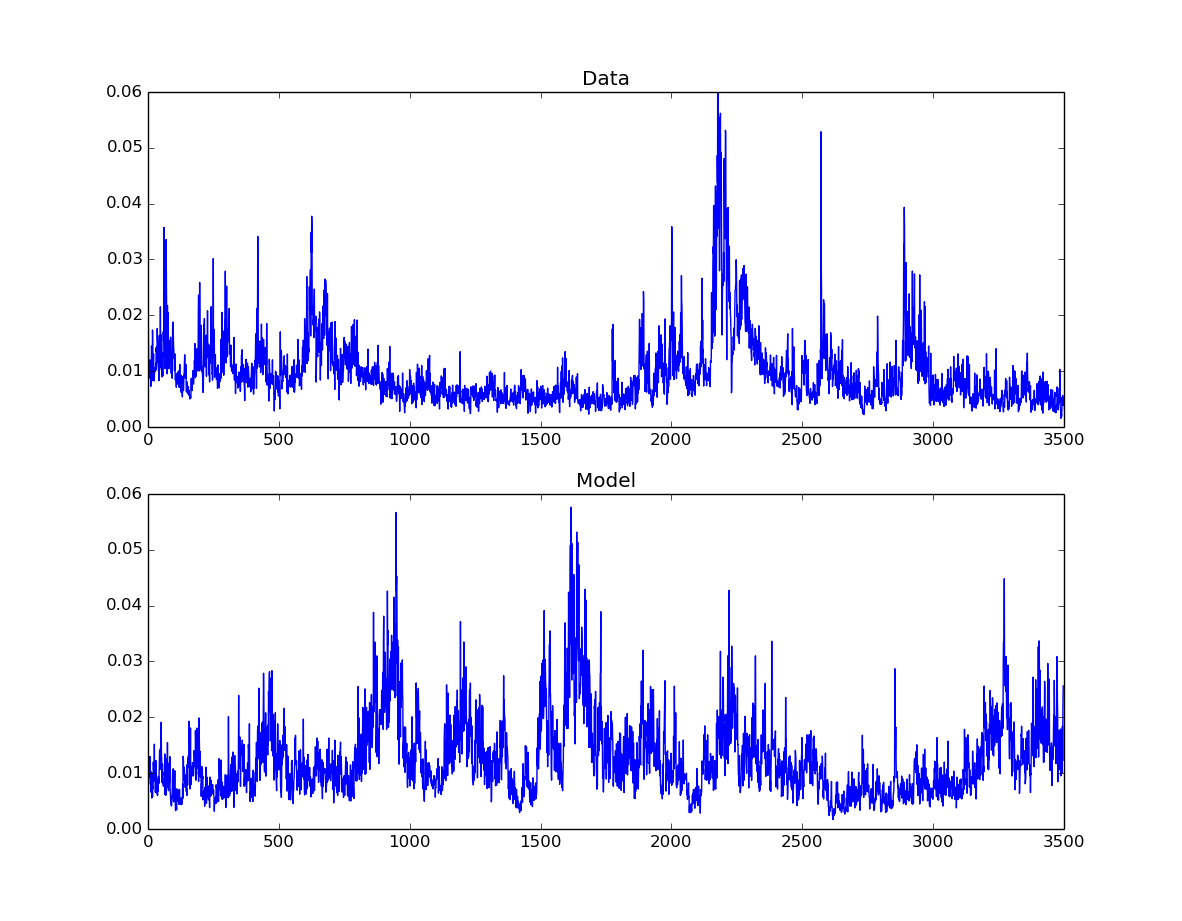}
\caption{Volatility of the S\&P (above) and of the model (below).}
\label{vol}
\end{center}
\end{figure}

\noindent We end this section by presenting in Figure \ref{vol} a sample path of the model-generated volatility (spot volatility direct from the simulation rather than estimated from the simulated price series) together with a graph of S\&P volatility over $3,500$ days. 

\noindent A first reaction to Figure \ref{vol} is that the simulated and actual graphs look very alike. In particular, in both of them, persistent periods of high volatility alternate with low volatility periods. On closer inspection of the empirical volatility series, we observe that the sample path of the volatility on a restricted time window seems to exhibit the same kind of qualitative properties as those of the global sample path (for example periods of high and low activity). This fractal-type behavior of the volatility has been investigated both empirically and theoretically in, for example, \cite{bacry2003log,bouchaud2003theory,mantegna2000introduction}.\\
 
\noindent At the visual level, we observe that this fractal-type behavior is also reproduced in our model, as we now explain. Denote by $L^{x,H}$ the law of the geometric fractional Brownian motion with Hurst exponent $H$ and volatility $x$ on $[0,1]$, that is $(e^{xW^H_t})_{t\in [0,1]}$. Then, when $\alpha$ is very small, the rescaled volatility process on $[0,\Delta]$: $(\sigma_{t\Delta}/\sigma_0)_{t\in [0,1]}$, has approximately the law $L^{\nu \Delta^H,H}$. Now remark that for $H$ small, the function $u^H$ increases very slowly. Thus, over a large range of observation scales $\Delta$, the rescaled volatility processes on $[0,\Delta]$ have approximately the same law. For example, between an observation scale of one day and five years (1250 open days), the coefficient $x$ characterizing the law of the volatility process is ``only'' multiplied by $1250^{0.14}=2.7$. It follows that in the RFSV model, the volatility process over one day resembles the volatility process over a decade.


\section{Spurious long memory of volatility?}
\label{statistics}
We revisit in this section the issue of long memory of volatility through the lens of our model. As mentioned earlier in the introduction, the long memory of volatility is widely accepted as a stylized fact. Specifically, this means that the autocovariance function $\text{Cov}[\text{log}(\sigma_t),\text{log}(\sigma_{t+\Delta})]$
(or sometimes $\text{Cov}[\sigma_t,\sigma_{t+\Delta}]$)
goes slowly to zero as $\Delta \to \infty$ and often even more precisely, that it behaves as $\Delta^{-\gamma}$, with $\gamma<1$ as $\Delta \to \infty$.\\


\noindent In previous sections, we showed that both in the data and in our model, 
$$
\text{Cov}[\text{log}(\sigma_t),\text{log}(\sigma_{t+\Delta})]\approx A-B\Delta^{2H}
$$
and 
$$
\text{Cov}[\sigma_t,\sigma_{t+\Delta}]\approx C\,e^{-B\Delta^{2H}}-D,
$$ 
for some constants $A$, $B$, $C$ and $D$.  Thus, neither in the model nor in the data does the autocovariance function decay as a power law. And neither the data nor the model exhibits long memory\footnote{In fact the notion of empirical long memory does not make much sense outside the power law case. Indeed the empirical values of covariances at very large time scales are never measurable and thus one cannot conclude if the series of covariances converges in general. All that we say here is that the autocovariance of the (log-)volatility does not behave as a power law.}, see again Figure \ref{log_cov_vol_log}.\\

\noindent We now revisit some standard statistical procedures aimed at identifying long memory that have been used in the financial econometrics literature. In the sequel, we apply these both to the data and to sample paths of the RFSV model.  Such procedures are of course designed to identify long memory under rather strict modeling assumptions;  spurious results may obviously then be obtained if the model underlying the estimation procedure is misspecified .\\

\begin{figure}[H]
\begin{center}
\includegraphics[width=9cm,height=8cm]{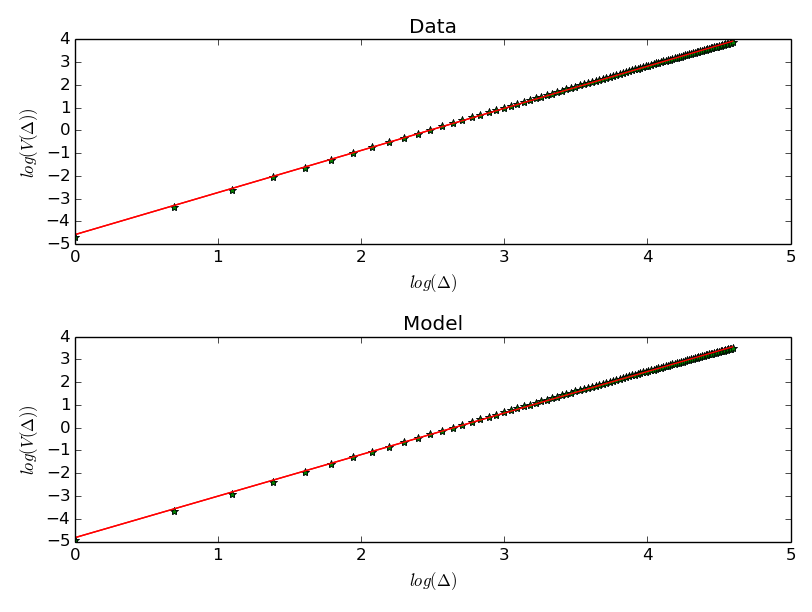}
\caption{Empirical counterpart of $\text{log}(V(t))$ as a function of $\text{log}(t)$ on S\&P (above) and simulation (below).}
\label{fig:andersen1}
\end{center}
\end{figure}

\noindent With the same model parameters as in Section \ref{sec:simu}, we simulate our model over 3,500 days, which corresponds to the size of our dataset. Consider first the procedure in \cite{andersen2001distribution}, where the authors test for long memory in the volatility by studying the scaling behavior of the quantity 
$$
V(t)=\text{Var}\left[\int_0^t \sigma_s^2 ds\right]
$$ 
with respect to $t$. In the model they consider, if $V(t)$ behaves asymptotically as $t^{2-\gamma}$ with $\gamma<1$, then the autocorrelation function of the log-volatility should behave as $t^{-\gamma}$. Figure \ref{fig:andersen1} presents the graph of the logarithm of the empirical counterpart of $V(t)$ against the logarithm of $t$, on the S\&P data and within our simulation framework.

\noindent We note from Figure \ref{fig:andersen1} that both our simulated model and market data lead to very similar graphs, close to straight lines with slope 1.86. Accordingly, in the setting of \cite{andersen2001distribution}, we would deduce power law behavior of the autocorrelation function with exponent $0.14$ and therefore long memory. Thus, if the data are generated by a model like the RFSV model, one can easily be wrongly convinced that the volatility time series exhibits long memory.\\ 

\begin{figure}[H]
\begin{center}
\includegraphics[width=9cm,height=8cm]{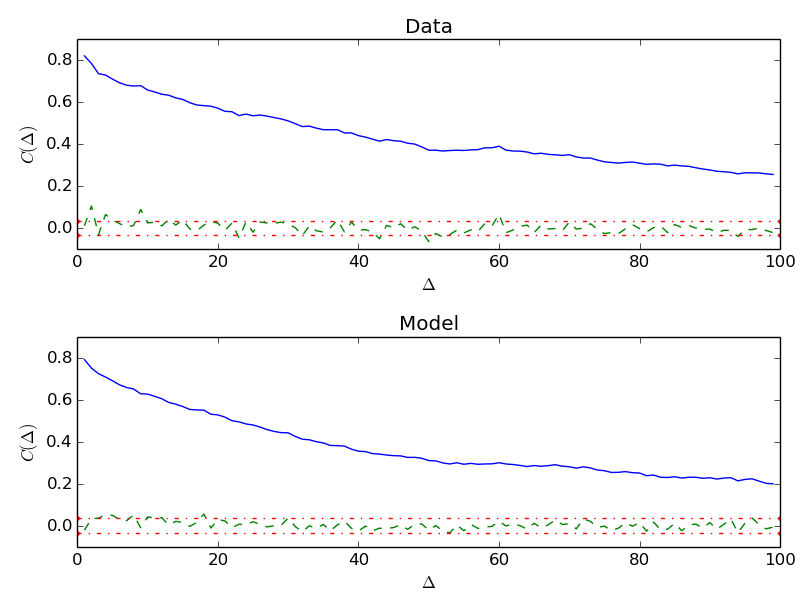}
\caption{Autocorrelation functions of $\text{log}(\sigma_t)$ (in blue) and $\varepsilon_t$ (in green) and the Bartlett standard error bands (in red), for S\&P data (above) and for simulated data (below).}
\label{andersen2}
\end{center}
\end{figure}

\noindent In \cite{andersen2003modeling}, the authors deduce long memory in the volatility by showing that the process $\varepsilon_t$ obtained by fractional differentiation of the log-volatility $\varepsilon_t=(1-L)^d\text{log}(\sigma_t)$, with $d=0.4$ (which is considered as a reasonable value) and $L$ the lag operator, behaves as a white noise. To check for this, they simply compute the autocorrelation function of $\varepsilon_t$. We give in Figure \ref{andersen2} the autocorrelation functions of the logarithm of $\sigma_t$ and $\varepsilon_t$, again both on the data and on the simulated path.

\noindent Once again, the data and the simulation generate very similar plots. We conclude that this procedure for estimating long memory is just as fragile as the first, and it is easy to wrongly deduce volatility long memory when applying it.\\

\noindent 
In conclusion, it seems that classical estimation  procedures identify spurious long memory of volatility in the RFSV model.  Moreover, these procedures estimate the same long memory parameter from data generated from a suitably calibrated RFSV model as they estimate from empirical data. Once again, our conclusion is that although the (log-)volatility may exhibit some form of persistence, it does not present any long memory in the classical power law sense.


\section{Forecasting using the RFSV model}
\label{varianceswaps}

In this section, we present an application of our model: forecasting the log-volatility and the variance.

\subsection{Forecasting log-volatility}\label{sec:logvolForecast}

The key formula on which our prediction method is based is the following one: 
$$
\E[W^H_{t+\Delta}|\cF_t]=\frac{\cos(H\pi)}{\pi} \Delta^{H+1/2} \int_{-\infty}^t \frac{W^H_s}{(t-s+\Delta)(t-s)^{H+1/2}} ds,
$$
where $W^H$ is a fBM with $H<1/2$ and $\cF_t$ the filtration it generates, see Theorem 4.2 of \cite{nuzman2000linear}. By construction, over any reasonable time scale of interest, as formalized in Corollary \ref{covfou}, we may approximate the fOU volatility process in the RFSV model as  $\log \sigma^2_{t} \approx 2\nu\,W^H_t+C$ for some constants $\nu$ and $C$. Our prediction formula for log-variance then follows:\footnote{The constants $2\nu$ and $C$ cancel when deriving the expression.}
\begin{equation}
\E\left[\log \sigma^2_{t+\Delta} | \cF_t\right] = \frac{\cos(H\pi)}{\pi} \Delta^{H+1/2} \int_{-\infty}^t \frac{\log \sigma_s^2}{(t-s+\Delta)(t-s)^{H+1/2}} ds.
\label{eq:RFSVforecast}
\end{equation}
This formula, or rather its approximation through a Riemann sum (we assume in this section that the volatilities are perfectly observed, although they are in fact estimated), is used to forecast the log-volatility 1, 5 and 20 days ahead $(\Delta=1,~5,~20)$.\\ 

\noindent We now compare the predictive power of formula \eqref{eq:RFSVforecast} with that of AR and HAR forecasts, in the spirit of \cite{corsi2009simple}\footnote{
Note that we do not consider GARCH models here since we have access to high frequency volatility estimates and not only to daily returns. Indeed, it is shown in \cite{andersen2003modeling} that forecasts based on the time series of realized variance outperform GARCH forecasts based on daily returns.}. Recall that for a given integer $p>0$, the AR(p) and HAR predictors take the following form (where the index $i$ runs over the series of daily volatility estimates):
\begin{itemize}
\item{AR(p):$$\widehat{\log(\sigma^2_{t+\Delta})}=K_0^{\Delta}+\sum_{i=0}^{p} C_i^{\Delta} \log(\sigma^2_{t-i}).$$}
\item{HAR :$$\widehat{\log(\sigma^2_{t+\Delta})}=K_0^{\Delta}+C_0^{\Delta}\log(\sigma^2_{t})+C_5^{\Delta} \frac{1}{5}\sum_{i=0}^{5}  \log(\sigma^2_{t-i})+C_{20}^{\Delta} \frac{1}{20}\sum_{i=0}^{20}  \log(\sigma^2_{t-i}).$$}
\end{itemize}
\noindent We estimate AR coefficients using the R \verb|stats| library\footnote{More precisely, we use the default Yule-Walker method.} on a rolling time window of $500$ days. In the HAR case, we use standard linear regression to estimate the coefficients as explained in \cite{corsi2009simple}. In the sequel, we consider $p=5$ and $p=10$ in the AR formula. Indeed, these parameters essentially give the best results for the horizons at which we wish to forecast the volatility (1, 5 and 20 days).
For each day, we forecast volatility for five different indices\footnote{In addition to S\&P and NASDAQ, we also investigate CAC40, FTSE and Nikkei, over the same time period as S\&P and NASDAQ. For simplicity, the parameter $H$ used in our predictor is computed only once for each asset, using the whole time period. This 
 yields similar results to using a moving time window adapted in time.}.\\

\noindent We then assess the quality of the various forecasts by computing the ratio $P$ between the mean squared error of our predictor and the (approximate) variance of the log-variance:
$$
P=\frac{\sum_{k=500}^{N-\Delta}\left(\log(\sigma^2_{k+\Delta})-\widehat{\log(\sigma^2_{k+\Delta})}\right)^2}{\sum_{k=500}^{N-\Delta} \left(\log(\sigma^2_{k+\Delta})-\E[\log(\sigma^2_{t+\Delta})]\right)^2},
$$
where $\E[\log(\sigma^2_{t+\Delta})]$ denotes the empirical mean of the log-variance over the whole time period.
\begin{table}[H]
\centering
\begin{tabular}{|c|c|c|c|c|}
\hline
     & AR(5) & AR(10) & HAR(3) & RFSV \\
\hline 
SPX2.rv  $\Delta=$ 1 & 0.317 & 0.318 & 0.314 & \textbf{ 0.313 } \\ 
\hline 
SPX2.rv  $\Delta=$ 5 & 0.459 & 0.449 & 0.437 & \textbf{ 0.426 } \\ 
\hline 
SPX2.rv  $\Delta=$ 20 & 0.764 & 0.694 & 0.656 & \textbf{ 0.606 } \\ 
\hline 
FTSE2.rv  $\Delta=$ 1 & 0.230 & 0.229 & 0.225 & \textbf{ 0.223 } \\ 
\hline 
FTSE2.rv  $\Delta=$ 5 & 0.357 & 0.344 & 0.337 & \textbf{ 0.320 } \\ 
\hline 
FTSE2.rv  $\Delta=$ 20 & 0.651 & 0.571 & 0.541 & \textbf{ 0.472 } \\ 
\hline 
N2252.rv  $\Delta=$ 1 & 0.357 & 0.358 & 0.351 & \textbf{ 0.345 } \\ 
\hline 
N2252.rv  $\Delta=$ 5 & 0.553 & 0.533 & 0.513 & \textbf{ 0.504 } \\ 
\hline 
N2252.rv  $\Delta=$ 20 & 0.875 & 0.795 & 0.746 & \textbf{ 0.714 } \\ 
\hline 
GDAXI2.rv  $\Delta=$ 1 & 0.237 & 0.238 & 0.234 & \textbf{ 0.231 } \\ 
\hline 
GDAXI2.rv  $\Delta=$ 5 & 0.372 & 0.362 & 0.350 & \textbf{ 0.339 } \\ 
\hline 
GDAXI2.rv  $\Delta=$ 20 & 0.661 & 0.590 & 0.550 & \textbf{ 0.498 } \\ 
\hline 
FCHI2.rv  $\Delta=$ 1 & 0.244 & 0.244 & 0.241 & \textbf{ 0.238 } \\ 
\hline 
FCHI2.rv  $\Delta=$ 5 & 0.378 & 0.373 & 0.366 & \textbf{ 0.350 } \\ 
\hline 
FCHI2.rv  $\Delta=$ 20 & 0.669 & 0.613 & 0.598 & \textbf{ 0.522 } \\ 
\hline
\end{tabular}
\caption{Ratio $P$ for the  AR, HAR and RFSV predictors.}
\label{tab:R2log}

\end{table}

\noindent We note from Table \ref{tab:R2log} that the RFSV forecast consistently outperforms the AR and HAR forecasts,  especially at longer horizons. Moreover, our forecasting method is more parsimonious since it only requires the parameter $H$ to forecast the log-variance. Compare this with the AR and HAR methods, for which coefficients depend on the forecast time horizon and must be recomputed if this horizon changes.\\


\noindent Remark that our predictor can be linked to that of \cite{duchon2012forecasting}, where the issue of the prediction of the log-volatility in the multifractal random walk model of \cite{bacry2003log} is tackled. In this model,
$$\E[\log(\sigma_{t+\Delta}^2)|\cF_t]=\frac{1}{\pi} \sqrt{\Delta} \int_{-\infty}^t \frac{\log(\sigma_s^2)}{(t-s+\Delta)\sqrt{t-s}} ds,$$
which is the limit of our predictor when $H$ tends to zero.\\

\noindent Note also that our prediction formula may be rewritten as
$$
\E[\log(\sigma^2_{t+\Delta})|\cF_t]=\frac{\cos(H\pi)}{\pi}\int_0^{+\infty} \frac{\log(\sigma_{t-\Delta u}^2)}{(u+1)\,u^{H+1/2}} du.
$$
For a given small $\varepsilon>0$, let $r$ be the smallest real number such that
$$
\int_{r}^{+\infty} \frac{1}{(u+1)\,u^{H+1/2}} du\leq \varepsilon.
$$
Then we have, with an error of order $\varepsilon$,
$$
\E[\log(\sigma^2_{t+\Delta})|\cF_t]\approx \frac{\cos(H\pi)}{\pi}\int_0^{r} \frac{\log(\sigma_{t-\Delta u}^2)}{(u+1)\,u^{H+1/2}} du.$$
Consequently, the volatility process needs to be considered (roughly) down to time $t-\Delta \,r$ if one wants to forecast up to time $\Delta$ in the future. The relevant regression window is thus linear in the forecasting horizon. For example, for $r=1$, $\varepsilon=0.35$ which is not so unreasonable.
In this case, as is well-known to practitioners, to predict volatility one week ahead, one should essentially look at the volatility over the last week. If trying to predict the volatility one month ahead, one should look at the volatility over the last month.

\subsection{Variance prediction}\label{sec:varForecast}

Recall that $\log \sigma_{t}^2 \approx 2\,\nu\,W^H_t+C$ for some constant $C$. In  \cite{nuzman2000linear}, it is shown that $W^H_{t+\Delta}$ is conditionally Gaussian with 
conditional variance
$$
\text{Var}[W^H_{t+\Delta}|\cF_t]=c\,\Delta^{2H}
$$
with $$
c = \frac{\Gamma(3/2-H)}{\Gamma(H+1/2)\,\Gamma(2-2H)}.
$$
Thus, we obtain the following natural form for the RFSV predictor of the variance:
$$
\widehat{\sigma^2_{t+\Delta}}=\exp\left\{\widehat{\log \sigma^2_{t+\Delta}}+2\, c\,\nu^2\Delta^{2\,H}\right\}
$$ 
where $\widehat{\log(\sigma^2_{t+\Delta})}$ is the estimator from Section \ref{sec:logvolForecast} and $\nu^2$ is estimated as the exponential of the intercept in the linear regression of $\log(m(2,\Delta))$ on $\log(\Delta)$.\\ 

\noindent As in the previous paragraph, we compare in Table \ref{R2} the performance of the RFSV forecast with those of AR and HAR forecasts (constructed on variance rather than log-variance this time). 
\begin{table}[H]
\begin{center}
\begin{tabular}{|c|c|c|c|c|}
	 \hline
     & AR(5) & AR(10) & HAR(3) & RFSV \\
\hline 
SPX2.rv  $\Delta=$ 1 & 0.520 & 0.566 & 0.489 & \textbf{ 0.475 } \\ 
\hline 
SPX2.rv  $\Delta=$ 5 & 0.750 & 0.745 & 0.723 & \textbf{ 0.672 } \\ 
\hline 
SPX2.rv  $\Delta=$ 20 & 1.070 & 1.010 & 1.036 & \textbf{ 0.903 } \\ 
\hline 
FTSE2.rv  $\Delta=$ 1 & 0.612 & 0.621 & 0.582 & \textbf{ 0.567 } \\ 
\hline 
FTSE2.rv  $\Delta=$ 5 & 0.797 & 0.770 & 0.756 & \textbf{ 0.707 } \\ 
\hline 
FTSE2.rv  $\Delta=$ 20 & 1.046 & 0.984 & 0.935 & \textbf{ 0.874 } \\ 
\hline 
N2252.rv  $\Delta=$ 1 & 0.554 & 0.579 & \textbf{ 0.504 } & 0.505 \\ 
\hline 
N2252.rv  $\Delta=$ 5 & 0.857 & 0.807 & 0.761 & \textbf{ 0.729 } \\ 
\hline 
N2252.rv  $\Delta=$ 20 & 1.097 & 1.046 & 1.011 & \textbf{ 0.964 } \\ 
\hline 
GDAXI2.rv  $\Delta=$ 1 & 0.439 & 0.448 & 0.399 & \textbf{ 0.386 } \\ 
\hline 
GDAXI2.rv  $\Delta=$ 5 & 0.675 & 0.650 & 0.616 & \textbf{ 0.566 } \\ 
\hline 
GDAXI2.rv  $\Delta=$ 20 & 0.931 & 0.850 & 0.816 & \textbf{ 0.746 } \\ 
\hline 
FCHI2.rv  $\Delta=$ 1 & 0.533 & 0.542 & 0.470 & \textbf{ 0.465 } \\ 
\hline 
FCHI2.rv  $\Delta=$ 5 & 0.705 & 0.707 & 0.691 & \textbf{ 0.631 } \\ 
\hline 
FCHI2.rv  $\Delta=$ 20 & 0.982 & 0.952 & 0.912 & \textbf{ 0.828 } \\ 
\hline
\end{tabular}
\caption{Ratio $P$ for the AR, HAR and RFSV predictors.}
\label{R2}
\end{center}
\end{table}
\noindent We find again that the RFSV forecast typically outperforms HAR and AR, although it is worth noting that the HAR forecast is already visibly superior to the AR forecast.

\section{The microstructural foundations of the irregularity of the volatility}\label{hawkes}

We gather in this section some ideas which may help to understand why the observed volatility appears so irregular. The starting point is the analysis of the order flow through Hawkes processes. These processes are extensions of Poisson processes where the intensity at a given time depends on the location of the past jumps. More precisely, let us consider a time period starting at $0$ and denote by $N_t$ the number of transactions between $0$ and $t$. Assuming the point process $N_t$ follows a Hawkes process means its intensity at time $t$, $\lambda_t$, takes the form:
$$\lambda_t=\mu+\sum_{0<J_i<t}\phi(t-J_i),$$
where the $J_i$ are the past jump times, $\mu$ is a positive constant and $\phi$ is a non negative deterministic function called kernel.\\

\noindent When trying to calibrate such models on high frequency data, two main phenomena almost systematically occur:

\begin{itemize}
\item The $L^1$ norm of $\phi$ is close to one, see \cite{filimonov2012quantifying,filimonov2013apparent,hardiman2013critical,lallouache2014statistically}.
\item The function $\phi$ has a power law tail, see \cite{bacry2014hawkes,hardiman2013critical}.
\end{itemize}

\noindent The first of these two facts means the degree of endogeneity of the market is very high, that is one given order endogenously generates many other orders, see \cite{filimonov2012quantifying,filimonov2013apparent,hardiman2013critical}. This recent feature of financial markets is obviously related to electronic high frequency trading, where market participants automatically react to other participants orders through their algorithms. The second observation tells us that generally, a given order influences other orders over a long time period. This is likely due to the splitting of large orders. Indeed, many orders are actually part of a metaorder whose full execution can take a large amount of time.\\

\noindent We believe these two phenomena together lead to a superposition effect inducing this irregular volatility. Indeed, it is explained in \cite{jaisson2013limit,jaisson2014fractional} that the macroscopic scaling limit of Hawkes processes with power law tail and kernel with $L^1$ norm close to one can be seen as an integrated fractional process, with Hurst parameter $H$ smaller than $1/2$. This signifies that at large sampling scales, the dynamics of the cumulated order flow is well approximated by an integrated fractional process, with $H<1/2$. Then, it is clearly established that there is a linear relation between cumulated order flow and integrated variance. Thus we retrieve here that because of this superposition effect, the volatility should behave as a fractional process with $H<1/2$.   


\section{Conclusion}

Using daily realized variance estimates as proxies for daily spot (squared) volatilities, we uncovered two startlingly simple regularities in the resulting time series.  First we found that the distributions of increments of log-volatility are approximately Gaussian, consistent with many prior studies.  Secondly, we established the monofractal scaling relationship 
\begin{equation}
\E\left[|\log(\sigma_{\Delta})-\log(\sigma_0)|^q\right]=K_q \,\nu^q \, \Delta^{q\,H},
\label{eq:monofractal}
\end{equation}
where $H$ can be seen as a measure of smoothness characteristic of the underlying volatility process; typically, $0.06 < H <0.2$. The simple scaling relationship \eqref{eq:monofractal} naturally suggests that log-volatility may be modeled using fractional Brownian motion.\\ 

\noindent The resulting Rough Fractional Stochastic Volatility (RFSV) model turns out to be formally almost identical to the FSV model of Comte and Renault \cite{comte1998long}, with one major difference: In the FSV model, $H>1/2$ to ensure long memory whereas in the RFSV model $H<1/2$, typically, $ H \approx 0.1$.  Moreover, in the FSV model, the mean reversion coefficient $\alpha$ has to be large compared to $1/T$ to ensure a decaying volatility skew; in the RFSV model, the volatility skew decays naturally just like the observed volatility skew, $\alpha \ll 1/T$ and indeed for time scales of practical interest, we may proceed as if $\alpha$ were exactly zero. \\

\noindent We further showed that applying standard statistical estimators to volatility time series simulated with the RFSV model would lead us to erroneously deduce the presence of long memory, with parameters similar to those found in prior studies.  Despite that volatility in the RFSV model (or in the data) is not long memory, we can therefore explain why long memory of volatility is widely accepted as a stylized fact.\\

\noindent As an application of the RFSV model, we showed how to forecast volatility at various times cales, at least as well as Fulvio Corsi's impressive HAR estimator, but with only one parameter -- $H$!\\

\noindent Finally, we explained how the RFSV model could emerge as the scaling limit of a Hawkes process description of order flow.\\

\noindent In future work, we will explore the implications of the RFSV model (written under the physical measure $\mathbb{P}$), for option pricing (under the pricing measure $\mathbb{Q}$).  In particular, following Mandelbrot and Van Ness, the fBM that appears in the definition \eqref{eq:RFSV} of the RFSV model may be represented as a fractional integral of a standard Brownian motion as follows \cite{mandelbrot1968fractional}:
\begin{equation}
W^H_t=\int_{0}^t \frac{dW_s}{(t-s)^\gamma}+\int_{-\infty}^0 \,\left[\frac{1}{(t-s)^\gamma}-\frac{1}{(-s)^\gamma}\right]\,dW_s,
\label{eq:MandelbrotVanNess}
\end{equation}
with $\gamma = \frac 12 -H$.  The observed anticorrelation between price moves and volatility moves may then be  modeled naturally by anticorrelating the Brownian motion $W$ that drives the volatility process with the Brownian motion driving the price process.  As already shown by Fukasawa \cite{fukasawa2011asymptotic},  such a model with a small $H$ reproduces the observed decay of at-the-money volatility skew with respect to time to expiry, asymptotically for short times.  We will show that an appropriate extension of Fukasawa's model, consistent with the RFSV model, fits the entire implied volatility surface remarkably well, not just for short expirations.  Moreover, despite that it would seem from \eqref{eq:MandelbrotVanNess} that knowledge of the entire path $\{W_s:s<t\}$ of the Brownian motion would be required, it turns out that the statistics of this path necessary for option pricing are traded and thus easily observed.

\appendix

\section{Proofs}
\subsection{Proof of Proposition \ref{theoOUfBM}}\label{sec:proof3.1}

Starting from Equation \eqref{eq:defOU} and applying integration by parts, we get
$$X^\alpha_t=\nu W^H_t-\int_{-\infty}^t\nu \alpha e^{-\alpha(t-s)}W^H_sds+m.$$
Therefore,
$$(X^\alpha_t-X^\alpha_0)-\nu W^H_t=-\int_0^t \nu\alpha e^{-\alpha(t-s)}W^H_sds-\int_{-\infty}^0 \nu\alpha (e^{-\alpha(t-s)}-e^{\alpha s})W^H_sds.$$
Consequently,
$$\sup_{t\in [0,T]}|(X^\alpha_t-X^\alpha_0)-\nu W^H_t|\leq \nu \alpha T \hat{W}^H_T +\int_{-\infty}^0 \nu \alpha (e^{\alpha s}-e^{-\alpha(T-s)})\hat{W}^H_sds,$$
where $\hat{W}^H_t=\sup_{s\in [0,t]}|W^H_s|$. Using the maximum inequality of \cite{novikov1999some}, we get
$$\E\big[\sup_{t\in [0,T]}|(X^\alpha_t-X^\alpha_0)-\nu W^H_t|\big]\leq c\big(\nu\alpha T T^H+\int_{-\infty}^0 \nu \alpha (T\alpha e^{ \alpha s}) |s|^Hds\big),$$
with $c$ some constant. The term on the right hand side is easily seen to go to zero as $\alpha$ tends to zero.

\subsection{Proof of Corollary \ref{covfou}}\label{sec:proofCor3.1}

We first recall Equation (2.2) in \cite{cheridito2003fractional} which writes:
$$\text{Cov}[X^\alpha_{t+\Delta},X^\alpha_{t}]=K\int_\mathbb{R} e^{i\Delta x}\frac{|x|^{1-2H}}{\alpha^2+x^2}dx,$$ with $K=\nu^2\Gamma(2H+1)\text{sin}(\pi H)/(2\pi)$\footnote{This covariance is real because it is the Fourier transform of an even function.}. Now remark that
$$\E[(X^\alpha_{t+\Delta}-X^\alpha_{t})^2]=2\text{Var}[X_t^{\alpha}]-2\text{Cov}[X^\alpha_{t+\Delta},X^\alpha_{t}].$$
Therefore,
$$\E[(X^\alpha_{t+\Delta}-X^\alpha_{t})^2]=2K\int_\mathbb{R} (1-e^{i\Delta x})\frac{|x|^{1-2H}}{\alpha^2+x^2}dx.$$
This implies that for fixed $\Delta$, $\E[|X^\alpha_{t+\Delta}-X^\alpha_{t}|^2]$ is uniformly bounded by $$2K\int_\mathbb{R} (1-e^{i\Delta x})\frac{|x|^{1-2H}}{x^2}dx.$$
Moreover, $X^\alpha_{t+\Delta}-X^\alpha_{t}$ is a Gaussian random variable and thus for every $q$, its $(q+1)^{th}$ moment is uniformly bounded (in $\alpha$) so that the family $|X^\alpha_{t+\Delta}-X^\alpha_{t}|^q$ is uniformly integrable. Therefore, since by Proposition \ref{theoOUfBM}, 
\begin{center}
$|X^\alpha_{t+\Delta}-X^\alpha_{t}|^q\rightarrow \nu^q |W^H_{t+\Delta}-W^H_{t}|^q,$ in law,
\end{center} we get the convergence of the sequence of expectations.

\section{Estimations of $H$}

\subsection{On different indices}

\begin{table}[H]
\centering
\begin{tabular}{|l|c|c|c|c|c|}
  \hline
Index & $\zeta_{0.5}/0.5$ &$\zeta_1$ &$\zeta_{1.5}/1.5$ &$\zeta_2/2$ &$\zeta_3/3$  \\ 
\hline
SPX2.rv  & 0.128 & 0.126 & 0.125 & 0.124 & 0.124 \\ 
 \hline 
FTSE2.rv  & 0.132 & 0.132 & 0.132 & 0.131 & 0.127 \\ 
 \hline 
N2252.rv  & 0.131 & 0.131 & 0.132 & 0.132 & 0.133 \\ 
 \hline 
GDAXI2.rv  & 0.141 & 0.139 & 0.138 & 0.136 & 0.132 \\ 
 \hline 
RUT2.rv  & 0.117 & 0.115 & 0.113 & 0.111 & 0.108 \\ 
 \hline 
AORD2.rv  & 0.072 & 0.073 & 0.074 & 0.075 & 0.077 \\ 
 \hline 
DJI2.rv  & 0.117 & 0.116 & 0.115 & 0.114 & 0.113 \\ 
 \hline 
IXIC2.rv  & 0.131 & 0.133 & 0.134 & 0.135 & 0.137 \\ 
 \hline 
FCHI2.rv  & 0.143 & 0.143 & 0.142 & 0.141 & 0.138 \\ 
 \hline 
HSI2.rv  & 0.079 & 0.079 & 0.079 & 0.080 & 0.082 \\ 
 \hline 
KS11.rv  & 0.133 & 0.133 & 0.134 & 0.134 & 0.132 \\ 
 \hline 
AEX.rv  & 0.145 & 0.147 & 0.149 & 0.149 & 0.149 \\ 
 \hline 
SSMI.rv  & 0.149 & 0.153 & 0.156 & 0.158 & 0.158 \\ 
 \hline 
IBEX2.rv  & 0.138 & 0.138 & 0.137 & 0.136 & 0.133 \\ 
 \hline 
NSEI.rv  & 0.119 & 0.117 & 0.114 & 0.111 & 0.102 \\ 
 \hline 
MXX.rv  & 0.077 & 0.077 & 0.076 & 0.075 & 0.071 \\ 
 \hline 
BVSP.rv  & 0.118 & 0.118 & 0.119 & 0.120 & 0.120 \\ 
 \hline 
GSPTSE.rv  & 0.106 & 0.104 & 0.103 & 0.102 & 0.101 \\ 
 \hline 
STOXX50E.rv  & 0.139 & 0.135 & 0.130 & 0.123 & 0.101 \\ 
 \hline 
FTSTI.rv  & 0.111 & 0.112 & 0.113 & 0.113 & 0.112 \\ 
 \hline 
FTSEMIB.rv  & 0.130 & 0.132 & 0.133 & 0.134 & 0.134 \\ 
 \hline 
\end{tabular}
 \caption{Estimates of $\zeta_q$ for all indices in the Oxford-Man dataset.}
\label{table:OxfordSummary}
\end{table}

\subsection{On different time intervals\footnote{Note that we used realized kernel rather than realized variance estimates to generate Table \ref{table:TimeVaryingH}.  Results obtained using different variance estimators are almost indistinguishable.}}

\begin{table}[H]
\centering
\begin{tabular}{|l|c|c|}
  \hline
 Index & H (first half) & H (second half) \\ 
  \hline
SPX2.rk & 0.115 & 0.158 \\
 \hline  
 FTSE2.rk & 0.140 & 0.156 \\ 
 \hline 
  N2252.rk & 0.083 & 0.134 \\ 
 \hline   
GDAXI2.rk & 0.154 & 0.168 \\ 
 \hline 
  RUT2.rk & 0.098 & 0.149 \\ 
 \hline 
  AORD2.rk & 0.059 & 0.114 \\ 
 \hline 
  DJI2.rk & 0.123 & 0.151 \\ 
 \hline 
  IXIC2.rk & 0.094 & 0.156 \\ 
 \hline 
  FCHI2.rk & 0.140 & 0.146 \\ 
 \hline 
  HSI2.rk & 0.072 & 0.129 \\ 
 \hline 
  KS11.rk & 0.109 & 0.147 \\ 
 \hline 
  AEX.rk & 0.168 & 0.151 \\ 
 \hline 
  SSMI.rk & 0.206 & 0.183 \\ 
 \hline 
  IBEX2.rk & 0.122 & 0.149 \\ 
 \hline 
  NSEI.rk & 0.112 & 0.124 \\ 
 \hline 
  MXX.rk & 0.068 & 0.118 \\ 
 \hline 
  BVSP.rk & 0.074 & 0.134 \\ 
 \hline 
  GSPTSE.rk & 0.075 & 0.147 \\ 
 \hline 
  STOXX50E.rk & 0.138 & 0.132 \\ 
 \hline 
  FTSTI.rk & 0.080 & 0.171 \\ 
 \hline 
  FTSEMIB.rk & 0.133 & 0.140 \\ 
   \hline
\end{tabular}
\caption{Estimates of $H$ over two different time intervals for all indices in the Oxford-Man dataset}
\label{table:TimeVaryingH}
\end{table}

\section{The effect of smoothing}\label{sec:fSS}

Although we are really interested in the model
\[
\log \sigma_{t+\Delta} - \log \sigma_t = \nu\,\left(W^H_{t+\Delta}-W_t^H\right),
\]
consider the more tractable (fractional Stein and Stein or fSS) model:
\[
 v_{t+\Delta} -v_t = \alpha\,\left(W^H_{t+\Delta}-W_t^H\right),
\]
where $v_t =  \sigma^2$.  We cannot observe $v_t$ but suppose we can proxy it by the average
\[
\hat v^\delta_t = \frac 1 \delta \int_0^\delta\,v_u\,du.
\]

\noindent We would, for example, like to estimate 
$
m(2,\Delta) = \E\left[  (v_{t+\Delta}-  v_{t})^2 \right]
$.
However, we need to proxy spot variance with integrated variance so instead we have the estimate
\begin{eqnarray}
m^\delta(2,\Delta) &=& \E\left[  (\hat v^\delta_{t+\Delta}-  \hat v^\delta_{t})^2 \right]\nonumber\\
&=& \frac{1}{\delta^2}\,\E\left[\left(\int_0^\delta\, (v_{u+\Delta}-v_u)\,du\right)^2\right]\nonumber\\
&=&\frac{\alpha^2}{\delta^2}\,\int_0^\delta\,\int_0^\delta\,\E\left[ (W^H_{u+\Delta}-W^H_u)\,(W^H_{s+\Delta}-W^H_s) \right]\,du\,ds\nonumber\\
&=&\int_0^\delta\,\int_0^\delta\,\left\{  |u-s+\Delta|^{2\,H}-|u-s|^{2\,H} \right\}\,du\,ds, 
\label{eq:m2Delta}
\end{eqnarray}
where the last step uses that:
\[
\E\left[W^H_u\,W^H_s\right] = \frac12\,\left\{u^{2\,H} + s^{2\,H} - |u-s|^{2\,H}\right\},
\]
and the symmetry of the integral.\\

\noindent We assume that the length $\delta$ of the smoothing window is less than one day so $\Delta > \delta$.  Then easy computations give
\begin{eqnarray*}
&&\int_0^\delta\,\int_0^\delta\, |u-s+\Delta|^{2\,H}\,du\,ds\\
&=& \frac{1}{2\,H+1}\, \frac{1}{2\,H+2}\,\left\{ (\Delta+\delta)^{2\,H+2} -2\, \Delta^{2\,H+2}
+(\Delta-\delta)^{2\,H+2}\right\}
\end{eqnarray*}
and 
\begin{eqnarray*}
\int_0^\delta\,\int_0^\delta\,|u-s|^{2\,H}\,du\,ds 
&=&\frac2{2\,H+1}\,\frac{1}{2\,H+2}\,\delta^{2\,H+2}.
\end{eqnarray*}
Substituting back into \eqref{eq:m2Delta} gives 
\begin{eqnarray*}
m^\delta(2,\Delta) &=& \alpha^2\,\Delta^{2\,H}\, \frac{1}{2\,H+1}\,\frac{1}{2\,H+2}\,
\frac{1}{\theta^2}\,\left\{ (1+\theta)^{2\,H+2} -2 - 2\,\theta^{2\,H+2}
+(1-\theta)^{2\,H+2}  \right\}\\
&=:&\alpha^2\,\Delta^{2\,H}\,f(\theta).
\end{eqnarray*}
where $\theta= \delta/\Delta$.\\

\begin{figure}[!tbh]
\begin{center}
\includegraphics[width=0.9\linewidth]{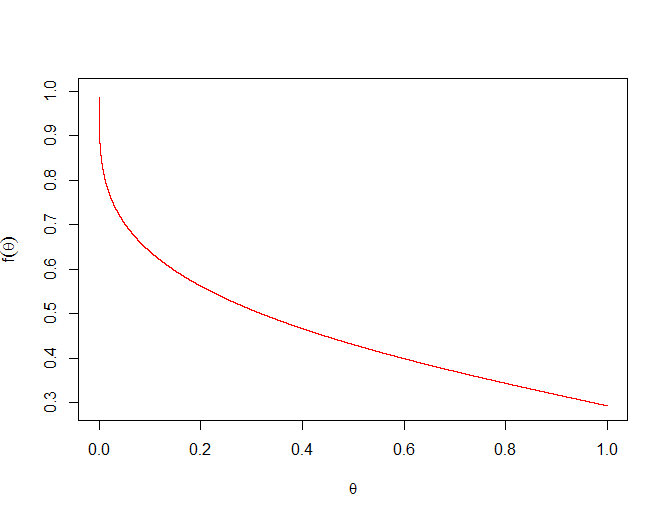}
\caption{$f(\theta)$ vs $\theta=\delta/\Delta$ with $H=0.14$.}
\label{fig:fTheta}
\end{center}
\end{figure}

\noindent Figure \ref{fig:fTheta} shows the effect of smoothing on the estimated variance in the fSS model.
Keeping $\delta$ fixed, as $\Delta$ increases, $f(\theta) = f(\delta/\Delta)$ increases towards one.  Thus, in a linear regression of $\log m^\delta(2,\Delta)$ against $\log \Delta$, we will obtain a higher effective $H$ (from the higher slope) and a lower effective (``volatility of volatility'') $\alpha$, exactly as we observed in the RSFV model simulations in Section \ref{sec:simu}. 

\subsubsection*{Numerical example}

\begin{figure}[!tbh]
\begin{center}
\includegraphics[width=\linewidth]{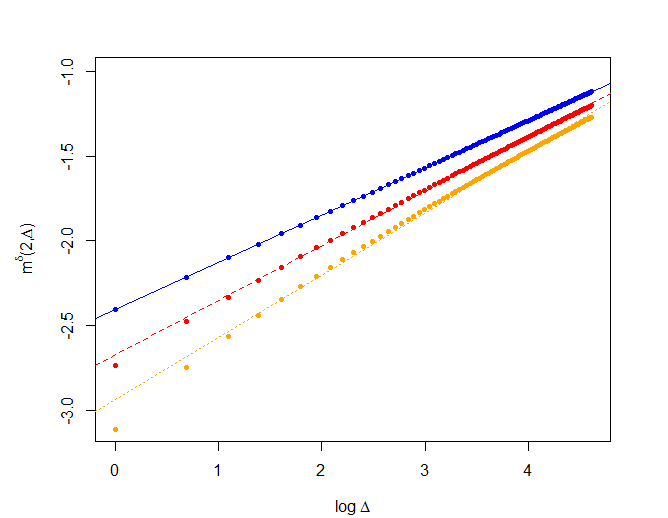}
\caption{Analogue of Figure \ref{scale_incr_log_vol_simu_1h-8h} in the fSS model:  The blue solid line is the true $m(2,\Delta)$; the red long-dashed line is the UZ estimate $m^{\delta_1}(2,\Delta)$; the orange short-dashed line is the RV estimate $m^{\delta_2}(2,\Delta)$.}
\label{fig:Regressions}
\end{center}
\end{figure}

In the simulation of the RSFV model in Section \ref{sec:simu}, we have $H=0.14$, $\delta_1 =1/24$ for the UZ estimate and $\delta_2 = 1/3$ for the RV estimate.  We now reproduce a fSS analogue of the RFSV simulation plots of $m(2,\Delta)$ in Figure \ref{scale_incr_log_vol_simu_1h-8h}.  Specifically, for each $\Delta \in \{1,2,...,100\}$, with $\alpha=0.3$  and $\delta=\delta_1$ or $\delta=\delta_2$, we compute the $m^\delta(2,\Delta)$ and regress $\log m^\delta(2,\Delta)$ against $\log \Delta$.  The regressions are shown in Figure \ref{fig:Regressions} and results tabulated in Table \ref{tab:regressions}.


\noindent In Figure \ref{fig:Regressions} and Table \ref{tab:regressions}, we observe similar qualitative and quantitative biases from our fSS model simulation as we observe in our simulation of the RSFV model with equivalent parameters in Section \ref{sec:simu}.

\begin{table}[!htdp]
\begin{center}
\begin{tabular}{lcc}
Estimate & Est. $\alpha$ &Est. $H$ \\
\hline
Exact ($\delta=0$) &      0.300 &       0.140 \\
 UZ ($\delta=1/24$) &  0.263 &   0.161\\ 
RV ($\delta=1/3$)&  0.230 &   0.184\\  
\hline
\end{tabular}
\end{center}
\caption{Estimated model parameters from the regressions shown in Figure \ref{fig:Regressions}.}
\label{tab:regressions}
\end{table}%

\bibliographystyle{abbrv}
\bibliography{bibli_VNLM}

\begin{thebibliography}{10}

\bibitem{andersen1997intraday}
T.~G. Andersen and T.~Bollerslev.
\newblock Intraday periodicity and volatility persistence in financial markets.
\newblock {\em Journal of Empirical Finance}, 4(2):115--158, 1997.

\bibitem{andersen2001stock}
T.~G. Andersen, T.~Bollerslev, F.~X. Diebold, and H.~Ebens.
\newblock The distribution of realized stock return volatility.
\newblock {\em Journal of Financial Economics}, 61(1):43--76, 2001.

\bibitem{andersen2001distribution}
T.~G. Andersen, T.~Bollerslev, F.~X. Diebold, and P.~Labys.
\newblock The distribution of realized exchange rate volatility.
\newblock {\em Journal of the American Statistical Association},
  96(453):42--55, 2001.

\bibitem{andersen2003modeling}
T.~G. Andersen, T.~Bollerslev, F.~X. Diebold, and P.~Labys.
\newblock Modeling and forecasting realized volatility.
\newblock {\em Econometrica}, 71(2):579--625, 2003.

\bibitem{bacry2003log}
E.~Bacry and J.~F. Muzy.
\newblock Log-infinitely divisible multifractal processes.
\newblock {\em Communications in Mathematical Physics}, 236(3):449--475, 2003.

\bibitem{bacry2014hawkes}
E.~Bacry and J.-F. Muzy.
\newblock Hawkes model for price and trades high-frequency dynamics.
\newblock {\em Quantitative Finance}, 14(7):1147--1166, 2014.

\bibitem{bentes2011stock}
S.~R. Bentes and M.~M. Cruz.
\newblock Is stock market volatility persistent? {A} fractionally integrated
  approach.
\newblock 2011.

\bibitem{beran1994statistics}
J.~Beran.
\newblock {\em Statistics for long-memory processes}, volume~61.
\newblock CRC Press, 1994.

\bibitem{bouchaud2003theory}
J.-P. Bouchaud and M.~Potters.
\newblock {\em Theory of financial risk and derivative pricing: {F}rom
  statistical physics to risk management}.
\newblock Cambridge University Press, 2003.

\bibitem{carr2003type}
P.~Carr and L.~Wu.
\newblock What type of process underlies options? {A} simple robust test.
\newblock {\em Journal of Finance}, 58(6):2581--2610, 2003.

\bibitem{chen2006persistence}
Z.~Chen, R.~T. Daigler, and A.~M. Parhizgari.
\newblock Persistence of volatility in futures markets.
\newblock {\em Journal of Futures Markets}, 26(6):571--594, 2006.

\bibitem{cheridito2003fractional}
P.~Cheridito, H.~Kawaguchi, and M.~Maejima.
\newblock Fractional {O}rnstein-{U}hlenbeck processes.
\newblock {\em Electron. J. Probab}, 8(3):14, 2003.

\bibitem{chronopoulou2011parameter}
A.~Chronopoulou.
\newblock Parameter estimation and calibration for long-memory stochastic
  volatility models.
\newblock In F.~G. Viens, M.~C. Mariani, and I.~Florescu, editors, {\em
  Handbook of Modeling High-Frequency Data in Finance}, pages 219--231. John
  Wiley \& Sons, 2011.

\bibitem{chronopoulou2012estimation}
A.~Chronopoulou and F.~G. Viens.
\newblock Estimation and pricing under long-memory stochastic volatility.
\newblock {\em Annals of Finance}, 8(2-3):379--403, 2012.

\bibitem{comte2012affine}
F.~Comte, L.~Coutin, and E.~Renault.
\newblock Affine fractional stochastic volatility models.
\newblock {\em Annals of Finance}, 8(2-3):337--378, 2012.

\bibitem{comte1998long}
F.~Comte and E.~Renault.
\newblock Long memory in continuous-time stochastic volatility models.
\newblock {\em Mathematical Finance}, 8(4):291--323, 1998.

\bibitem{Cont2007}
R.~Cont.
\newblock Volatility clustering in financial markets: Empirical facts and
  agent-based models.
\newblock In G.~Teyssi\`ere and A.~P. Kirman, editors, {\em Long Memory in
  Economics}, pages 289--309. Springer Berlin Heidelberg, 2007.

\bibitem{corsi2009simple}
F.~Corsi.
\newblock A simple approximate long-memory model of realized volatility.
\newblock {\em Journal of Financial Econometrics}, 7(2):174--196, 2009.

\bibitem{dayri2013large}
K.~Dayri and M.~Rosenbaum.
\newblock Large tick assets: Implicit spread and optimal tick size.
\newblock {\em Working paper}, 2013.

\bibitem{ding1993long}
Z.~Ding, C.~W. Granger, and R.~F. Engle.
\newblock A long memory property of stock market returns and a new model.
\newblock {\em Journal of Empirical Finance}, 1(1):83--106, 1993.

\bibitem{duchon2012forecasting}
J.~Duchon, R.~Robert, and V.~Vargas.
\newblock Forecasting volatility with the multifractal random walk model.
\newblock {\em Mathematical Finance}, 22(1):83--108, 2012.

\bibitem{dupire1994pricing}
B.~Dupire.
\newblock Pricing with a smile.
\newblock {\em Risk Magazine}, 7(1):18--20, 1994.

\bibitem{filimonov2012quantifying}
V.~Filimonov and D.~Sornette.
\newblock Quantifying reflexivity in financial markets: {T}oward a prediction
  of flash crashes.
\newblock {\em Physical Review E}, 85(5):056108, 2012.

\bibitem{filimonov2013apparent}
V.~Filimonov and D.~Sornette.
\newblock Apparent criticality and calibration issues in the {H}awkes
  self-excited point process model: {A}pplication to high-frequency financial
  data.
\newblock {\em arXiv preprint arXiv:1308.6756}, 2013.

\bibitem{fukasawa2011asymptotic}
M.~Fukasawa.
\newblock Asymptotic analysis for stochastic volatility: {M}artingale
  expansion.
\newblock {\em Finance and Stochastics}, 15(4):635--654, 2011.

\bibitem{gatheral2006volatility}
J.~Gatheral.
\newblock {\em The volatility surface: {A} practitioner's guide}, volume 357.
\newblock John Wiley \& Sons, 2006.

\bibitem{gatheral2014arbitrage}
J.~Gatheral and A.~Jacquier.
\newblock Arbitrage-free {SVI} volatility surfaces.
\newblock {\em Quantitative Finance}, 14(1):59--71, 2014.

\bibitem{gatheral2010zero}
J.~Gatheral and R.~C. Oomen.
\newblock Zero-intelligence realized variance estimation.
\newblock {\em Finance and Stochastics}, 14(2):249--283, 2010.

\bibitem{hagan2002managing}
P.~S. Hagan, D.~Kumar, A.~S. Lesniewski, and D.~E. Woodward.
\newblock Managing smile risk.
\newblock {\em Wilmott Magazine}, pages 84--108, 2002.

\bibitem{hardiman2013critical}
S.~J. Hardiman, N.~Bercot, and J.-P. Bouchaud.
\newblock Critical reflexivity in financial markets: {A} {H}awkes process
  analysis.
\newblock {\em arXiv preprint arXiv:1302.1405}, 2013.

\bibitem{heston1993closed}
S.~L. Heston.
\newblock A closed-form solution for options with stochastic volatility with
  applications to bond and currency options.
\newblock {\em Review of Financial Studies}, 6(2):327--343, 1993.

\bibitem{hull1993one}
J.~Hull and A.~White.
\newblock One-factor interest-rate models and the valuation of interest-rate
  derivative securities.
\newblock {\em Journal of Financial and Quantitative Analysis},
  28(02):235--254, 1993.

\bibitem{jaisson2013limit}
T.~Jaisson and M.~Rosenbaum.
\newblock Limit theorems for nearly unstable {H}awkes processes.
\newblock {\em The Annals of Applied Probability, to appear}, 2013.

\bibitem{jaisson2014fractional}
T.~Jaisson and M.~Rosenbaum.
\newblock Fractional diffusions as scaling limits of nearly unstable
  heavy-tailed {H}awkes processes.
\newblock {\em Working paper}, 2014.

\bibitem{lallouache2014statistically}
M.~Lallouache and D.~Challet.
\newblock Statistically significant fits of {H}awkes processes to financial
  data.
\newblock {\em Available at SSRN 2450101}, 2014.

\bibitem{mandelbrot1968fractional}
B.~B. Mandelbrot and J.~W. Van~Ness.
\newblock Fractional {B}rownian motions, fractional noises and applications.
\newblock {\em SIAM review}, 10(4):422--437, 1968.

\bibitem{mantegna2000introduction}
R.~N. Mantegna and H.~E. Stanley.
\newblock {\em Introduction to econophysics: {C}orrelations and complexity in
  finance}.
\newblock Cambridge University Press, 2000.

\bibitem{mikosch2000really}
T.~Mikosch and C.~St{\u{a}}ric{\u{a}}.
\newblock Is it really long memory we see in financial returns.
\newblock In P.~Embrechts, editor, {\em Extremes and integrated risk
  management}, pages 149--168. Risk Books, 2000.

\bibitem{musiela2006martingale}
M.~Musiela and M.~Rutkowski.
\newblock {\em Martingale methods in financial modelling}, volume~36.
\newblock Springer, 2006.

\bibitem{novikov1999some}
A.~Novikov and E.~Valkeila.
\newblock On some maximal inequalities for fractional {B}rownian motions.
\newblock {\em Statistics \& Probability Letters}, 44(1):47--54, 1999.

\bibitem{nuzman2000linear}
C.~J. Nuzman and V.~H. Poor.
\newblock Linear estimation of self-similar processes via {L}amperti's
  transformation.
\newblock {\em Journal of Applied Probability}, 37(2):429--452, 2000.

\bibitem{robert2011new}
C.~Y. Robert and M.~Rosenbaum.
\newblock A new approach for the dynamics of ultra-high-frequency data: {T}he
  model with uncertainty zones.
\newblock {\em Journal of Financial Econometrics}, 9(2):344--366, 2011.

\bibitem{robert2012volatility}
C.~Y. Robert and M.~Rosenbaum.
\newblock Volatility and covariation estimation when microstructure noise and
  trading times are endogenous.
\newblock {\em Mathematical Finance}, 22(1):133--164, 2012.

\bibitem{rosenbaum2008estimation}
M.~Rosenbaum.
\newblock Estimation of the volatility persistence in a discretely observed
  diffusion model.
\newblock {\em Stochastic Processes and their Applications}, 118(8):1434--1462,
  2008.

\bibitem{rosenbaum2009first}
M.~Rosenbaum.
\newblock First order p-variations and {B}esov spaces.
\newblock {\em Statistics \& Probability Letters}, 79(1):55--62, 2009.

\bibitem{rosenbaum2011new}
M.~Rosenbaum.
\newblock A new microstructure noise index.
\newblock {\em Quantitative Finance}, 11(6):883--899, 2011.

\end{thebibliography}

\end{document}